\documentclass[12pt,a4paper]{article}
\usepackage{jheppub}
\usepackage{amsmath,amssymb}
\usepackage{mathrsfs}
\usepackage[vcentermath,stdtext]{youngtab}
\usepackage{graphicx}
\usepackage{caption}
\usepackage{subcaption}
\usepackage{slashbox,booktabs}
\graphicspath{{fig/}}
\usepackage{hyperref}
\usepackage[active]{srcltx}

\DeclareMathOperator{\tr}{tr}
\DeclareMathOperator{\str}{str}

\DeclareMathOperator{\li}{Li}
\DeclareMathOperator{\pf}{pf}

\title{Wilson loops in \(\mathcal{N}=6\) superspace for ABJM theory}

\author[a]{M.~Rosso,}
\author[a]{C.~Vergu}

\affiliation[a]{ETH Z\"urich, Institut f\"ur Theoretische Physik,\\
Wolfgang Pauli Strasse 27,\\
CH-8093 Z\"urich, Switzerland}
\emailAdd{mrosso@itp.phys.ethz.ch}
\emailAdd{verguc@itp.phys.ethz.ch}

\abstract{In this paper we construct a light-like polygonal Wilson
  loop in \(\mathcal{N}=6\) superspace for ABJM theory.  We then use
  it to obtain constraints on its two- and three-loop bosonic version,
  by focusing on higher order terms in the \(\theta\) expansion.  The
  Grassmann expansion of the three-loop answer contains integrals
  which may be elliptic polylogarithms.  Our results take their
  simplest form when expressed in terms of \(OSp(6\vert 4)\)
  supertwistors.}

\begin{document}

\maketitle

\section{Introduction}
\label{sec:intro}

The fact that a certain operator or observable can be supersymmetrized
often has important consequences.  Even for an observable which is not
supersymmetric but admits a supersymmetric completion it can be useful
to consider it.  An early example of the usefulness of supersymmetry
for a non-supersymmetric quantity is the argument of
ref.~\cite{Grisaru1977} which uses supersymmetry to show that some
tree-level helicity amplitudes in Yang-Mills theory vanish.  Then Parke and
Taylor used supersymmetry to simplify the computations of some
non-vanishing helicity amplitudes (see refs.~\cite{Parke1985,
  Parke1986, PhysRevLett.56.2459}).

The observables we will consider in this paper are light-like
polygonal Wilson loops in ABJM theory~\cite{Aharony2008} (see also
refs.~\cite{doi:10.1142/S0217751X93001363, Schwarz:2004yj,
  Gaiotto2007, Bagger:2006sk, Bagger:2007jr, Gustavsson:2007vu,
  VanRaamsdonk:2008ft} for earlier related work).  The ABJM
theory is a theory of two Chern-Simons gauge connections interacting
with fermions and scalars in bifundamental representations.  It is a
superconformal theory with symmetry group \(OSp(6 \vert 4)\).  It is
therefore natural to build observables which are invariant under this superconformal symmetry.

The shape of a general Wilson loop changes in a complicated way under
conformal transformations.  The shape of a polygonal light-like
contour transforms in a much simpler way which makes it more
attractive to study (see however ref.~\cite{Muller2013}, where a
general Wilson loop shape is considered).

In pure Chern-Simons theory, appropriately regularized (framed) Wilson
loops yield topological invariants.  When adding matter fields, like
in the ABJM theory, the Wilson loops start depending on the shape of
their contour in a more complicated way.  In Yang-Mills theory, the
Wilson loop require a non-trivial renormalization as explained in
refs.~\cite{Polyakov1980, Dotsenko1980} for the Euclidean signature
and later in ref.~\cite{Korchemskaya1992} for Lorentzian signature.
If the contour of a Wilson loop in Chern-Simons theory has continuous
third derivatives, the one-loop perturbative correction is finite (see
for example ref.~\cite{Guadagnini1990}) but not topological invariant.

When introducing singularities, and just like their Yang-Mills counterpart, the
ABJM light-like Wilson loops become UV-divergent and these UV divergences break
the conformal symmetry (see ref.~\cite{Henn:2010ps}\footnote{Dimensional
  regularization in Chern--Simons is subtle, ee~\cite{Bianchi2013} for a
  detailed analysis.}).  In the following we will focus on
the finite parts.  As a rule, the UV-divergent terms, after regularization,
depend on the kinematics in relatively simple ways.  Therefore, by focusing on
the parts which are not UV divergent we capture the richest dependence on the
kinematics.

After supersymmetrization a further complication appears: the
right notion of light-like line is not one-dimensional anymore, but it
has one bosonic dimension and several fermionic dimensions (see
ref.~\cite{Witten1978}).  In fact, the equations of motion of some
gauge theories with extended supersymmetry are equivalent to the
flatness of the field strength on such light-like lines.  Despite
these complications a super-Wilson loop can be constructed following
the same steps as in ref.~\cite{Beisert2012a} and we do so in sec.~\ref{sec:super-wilson-loops}.

In four dimensions, for \(\mathcal{N}=4\) super-Yang-Mills, a similar
construction was done in chiral superspace in refs.~\cite{Mason2010,
  Caron-Huot2011}.  This construction is remarkable because of the
duality with scattering amplitudes, which was uncovered in a series of
papers~\cite{Alday2007a, Drummond:2007aua, Brandhuber2008,
  Drummond2008b, Bern2008a, Drummond:2008aq, Drummond:2008vq,
  Berkovits:2008ic, Kosower2011a}.

Later, in ref.~\cite{Caron-Huot2011a}, Caron-Huot considered a
\emph{nonchiral} version of the super-Wilson loop, but he only
considered it to the lowest order in the non-chiral variables
\(\bar{\theta}\).  By computing the terms of type \(\theta_{i}
\bar{\theta}_{j}\) where the \(\theta_{i},\bar{\theta}_{j}\)
correspond to different vertices\footnote{In fact, it turns out to be
  better to consider odd variables associated to the sides instead.}
of the  super-Wilson loop, he was able to compute the symbol
(see~\cite{Goncharov:2010jf}) of the two-loop planar \(n\)-sided
bosonic Wilson loop.

It is our objective in this paper to render this procedure more systematic on
the example of ABJM super-Wilson loops. Our focus is on the construction of null
polygonal supersymmetric Wilson loops; however, in principle there should be no
obstruction to defining a smooth supersymmetric Wilson loop, as done in
ref.~\cite{Muller2013} for \(\mathcal{N}=4\) super Yang--Mills.

\section{General discussion}
\label{sec:discussion}

We consider an $\mathcal{N} = 6$ Chern-Simons theory in three dimensions,
coupled to matter, with gauge covariant derivatives:
\begin{equation}
  \nabla_{\mu} = \partial_{\mu} + A_{\mu}, \qquad
  \nabla_{I J, \alpha} = D_{I J, \alpha} + A_{I J, \alpha},
\end{equation} where $\alpha$ is a spinor index transforming in the
two-dimensional spin representation of the three-dimensional Lorentz group
\(SO(1,2) \simeq SL(2, \mathbb{R})\) and the indices $I, J$ take four values and
transform in the representation \(\overline{\mathbf{4}}\) of \(SU(4)\)
when in the lower position and in \(\mathbf{4}\) when in the upper
position.  The supersymmetry covariant derivative \(D_{IJ}\) is
antisymmetric in the pair of indices \(IJ\) and is given by
\begin{equation}
  D_{I J, \alpha} = \frac \partial {\partial \theta^{I J, \alpha}} + (\gamma^{\mu} \bar{\theta}_{I J})_{\alpha} \partial_{\mu},
\end{equation} where \(\theta^{IJ}\) are the odd coordinates of
\(\mathcal{N}=6\) superspace.   Here the covariant derivatives and
gauge connections transform as $(\mathbf{Ad}, \mathbf{Ad})$ under
$U(N) \times U(M)$ gauge group.  They can be thought of as \((N+M)
\times (N+M)\) matrices with vanishing off-diagonal \(N \times M\) and
\(M \times N\) blocks.  The \(\bar{\theta}\) are not independent on
\(\theta\) but are related by \(\bar{\theta}_{IJ} = \tfrac 1 2
\epsilon_{IJKL} \theta^{KL}\).

Using
\begin{equation}
  \label{eq:ferm-derivative}
  \partial_{\alpha, I J} \bar{\theta}_{K L}^{\beta} = \frac 1 2 \epsilon_{K L M N} \partial_{\alpha, I J} \theta^{\beta, M N} = \frac 1 2 \epsilon_{K L M N} \delta_{\alpha}^{\beta} (\delta_{I}^{M} \delta_{J}^{N} - \delta_{J}^{M} \delta_{I}^{N}) = \epsilon_{I J K L} \delta_{\alpha}^{\beta},
\end{equation} we find
\begin{equation}
  \lbrace D_{I J, \alpha}, D_{K L, \beta}\rbrace = 2 \epsilon_{I J K L} \gamma^{\mu}_{\alpha \beta} \partial_{\mu}.
\end{equation}  From now on we will use the notation $\partial_{\alpha
  \beta} \equiv \gamma_{\alpha \beta}^{\mu} \partial_{\mu}$.  Under
hermitian conjugation we have \(\partial^{\dagger} = -\partial\),
\(D_{IJ}^{\dagger} = \frac 1 2 \epsilon^{IJKL} D_{KL}\).  The
gauge covariant derivatives have the same hermitian conjugation
properties as the supersymmetry covariant derivatives so we obtain
\(A_{\mu}^{\dagger} = -A_{\mu}\), \((A_{IJ})^{\dagger} = \frac 1 2
\epsilon^{IJKL} A_{KL}\).  We use conventions such that for two
Grassmann variables \(\psi\), \(\chi\) we have \((\psi \chi)^{\dagger}
= -\chi^{\dagger} \psi^{\dagger}\) and similarly for the transposition.  This explains the absence of some
factors of \(i\) which often appear in the literature on
supersymmetry.

The analysis of the superspace constraints for the $\mathcal{N} = 6$ and $\mathcal{N} = 8$ super-Chern-Simons theories has been done in refs.~\cite{Samtleben:2009ts, Samtleben:2010eu}.  The gauge covariant derivatives satisfy the following algebra
\begin{align}
  \lbrace \nabla_{I J, \alpha}, \nabla_{K L, \beta}\rbrace - 2 \epsilon_{I J K L} \nabla_{\alpha \beta} &= \epsilon_{\alpha \beta} \epsilon_{M I J [K} W^{M}_{\hphantom{M} L]} \equiv F_{I J \alpha, K L \beta},\\
  [\nabla_{I J, \alpha}, \nabla_{\beta \gamma}] &\equiv F_{I J \alpha, \beta \gamma},\\
  [\nabla_{\alpha \beta}, \nabla_{\gamma \delta}] &\equiv F_{\alpha \beta, \gamma \delta}.
\end{align}

In the first equation above we have already imposed some constraints
(the discussion follows closely the refs.~\cite{Samtleben:2009ts,
  Samtleben:2010eu} but with some change in conventions).  Let us
study the representations under the Lorentz \(SL(2)\) and
\(R\)-symmetry \(SU(4)\) groups.
The LHS transforms as $(\mathbf{2}, \mathbf{6}) \otimes_{S}
(\mathbf{2}, \mathbf{6}) = (\mathbf{3}, \mathbf{1}) \oplus
(\mathbf{1}, \mathbf{15}) \oplus (\mathbf{3}, \mathbf{20}')$ under
$SL(2) \times SU(4)$.\footnote{There are several conventions for
  naming the $SU(4)$ representations.  Here we use the conventions
  $\mathbf{20}' \simeq \yng(2,2)$, $\mathbf{20} \simeq \yng(2,2,1)$,
  $\overline{\mathbf{20}} \simeq \yng(2,1)$.}  We have already imposed
the conventional constraint that the $(\mathbf{3}, \mathbf{20}')$ part
vanishes.  According to the analysis in refs.~\cite{Samtleben:2009ts,
  Samtleben:2010eu} the inclusion of terms transforming as
$(\mathbf{3}, \mathbf{20}')$ is only needed when studying higher
derivative corrections.  The traceless field $W^{I}_{\hphantom{I} J}$
transforms as $\mathbf{15}$ of $SU(4)$.  Under hermitian conjugation
it transforms as \((W^{I}_{\hphantom{I} J})^{\dagger} =
W^{J}_{\hphantom{J} I}\).  Later we will show that it is not an independent field, but it is a composite of the matter fields charged under the Chern-Simons gauge group.  Another way to write the tensor $W$ is as $W_{I J, K L}$, which is antisymmetric in $(I, J)$, $(K, L)$ and in the exchange of the pairs $(I, J)$ and $(K, L)$.  The relation to the previous form is given by $W^{P}_{\hphantom{P} L} = - \tfrac 1 2 \epsilon^{I J K P} W_{I J, K L}$.

The Bianchi identity involving three covariant derivatives $\nabla_{I J \alpha}$, $\nabla_{K L \beta}$ and $\nabla_{M N \gamma}$ yields
\begin{equation}
  \epsilon_{K L M N} F_{I J \alpha, \beta \gamma} + \frac 1 2 \epsilon_{\beta \gamma} \epsilon_{P K L [M\vert} [\nabla_{I J \alpha}, W^{P}_{\hphantom{P} \vert N]}] + \text{cyclic} \lbrace (I J \alpha), (K L \beta), (M N \gamma)\rbrace = 0.
\end{equation}  This imposes some constraints on the derivatives of the $W$ superfield.  In order to extract these constraints let us contract this equation with $\epsilon^{K L M N}$.  After the contraction we obtain
\begin{equation}
  24 F_{I J \alpha, \beta \gamma} + 4 F_{I J \gamma, \alpha \beta} + 4 F_{I J \beta, \gamma \alpha} + \epsilon_{\alpha \beta} \times (\cdots)_{\gamma} + \epsilon_{\beta \gamma} \times (\cdots)_{\alpha} + \epsilon_{\gamma \alpha} \times (\cdots)_{\beta} = 0,
\end{equation} where we have indicated some of the terms only schematically.  This implies that the completely symmetric part (or spin $\tfrac 3 2$ under $SL(2)$) vanishes separately, that is
\begin{equation}
  F_{I J \alpha, \beta \gamma} + F_{I J \gamma, \alpha \beta} + F_{I J \beta, \gamma \alpha} = 0,
\end{equation} which can be solved as
\begin{equation}
  F_{I J \alpha, \beta \gamma} = \frac 1 3 \left(\epsilon_{\alpha \beta} F_{I J}^{\delta}{}_{, \gamma \delta} + \epsilon_{\alpha \gamma} F_{I J}^{\delta}{}_{, \beta \delta}\right).
\end{equation}

If after contracting with $\epsilon^{K L M N}$ we further contract with $\epsilon^{\alpha \beta}$, we finally obtain
\begin{equation}
  F_{I J}^{\hphantom{I J} \alpha}{}_{\gamma \alpha} = \frac 3 5 [\nabla_{P [I\vert \gamma}, W^{P}_{\hphantom{P} \vert J]}].
\end{equation}  From this one can reconstruct the field strength $F_{I J \alpha, \beta \gamma}$.  Plugging back into the Bianchi identity we find a constraint uniquely between the derivatives $\nabla_{I J} W^{K}_{\hphantom{K} L}$.  These quantities transform as $\mathbf{6} \otimes \mathbf{15} = \mathbf{6} \oplus \mathbf{10} \oplus \overline{\mathbf{10}} \oplus \mathbf{64}$ under $SU(4)$.  The constraint mentioned above forces the $\mathbf{64}$ part to vanish.  Therefore, the derivative of $W$ has a special form
\begin{equation}
  [\nabla_{I J \alpha}, W^{K}_{\hphantom{K} L}] = \delta_{[I\vert}^{K} \lambda_{\alpha \vert J] L} + \frac 1 4 \delta_{L}^{K} \lambda_{\alpha I J} + \delta_{[I\vert}^{K} \rho_{\alpha \vert J] L} - \frac 1 2 \epsilon_{I J L N} \bar{\rho}_{\alpha}^{K N},
\end{equation} where $\lambda_{I J}$ transforms in the $\mathbf{6}$ of $SU(4)$ while $\rho_{I J}$, $\bar{\rho}^{I J}$ transform in the $\mathbf{10}$ and $\overline{\mathbf{10}}$ respectively.  The form for the derivative of $W$ is fixed by the $SU(4)$ transformations, the tracelessness of $W$ and reality conditions.

Using the special form of $[\nabla, W]$ we can easily compute
\begin{equation}
  [\nabla_{P [I\vert \alpha}, W^{P}_{\hphantom{P} \vert J]}] = \frac 5 4 \lambda_{\alpha I J}, \qquad
  [\nabla_{P (I\vert \alpha}, W^{P}_{\hphantom{P} \vert J)}] = \frac 3 2 \rho_{\alpha I J}.
\end{equation}  Also,
\begin{equation}
  F_{I J \alpha, \beta \gamma} = \frac 1 4 \epsilon_{\alpha \beta} \lambda_{\gamma I J} + \frac 1 4 \epsilon_{\alpha \gamma} \lambda_{\beta I J}.
\end{equation}

The second Bianchi identity, between derivatives $\nabla_{\alpha \beta}$, $\nabla_{I J \gamma}$ and $\nabla_{K L \delta}$ gives the field strength $F_{\alpha \beta} \equiv \epsilon^{\gamma \delta} F_{\alpha \gamma, \beta \delta}$ in terms of covariant derivatives of $W$.  This is in fact the equation of motion for $F$ and we will return to it later, after we express $W$ in terms of fundamental fields.

It can be shown that the remaining Bianchi identities do not impose extra conditions.  Therefore, we have finished the analysis of the gauge sector and we now move on to analyzing the matter sector.  The matter fields are organized in two kinds of superfields: scalar $\Phi^{I}$ transforming in $\mathbf{4}$ of $SU(4)$ and $\Psi_{I \alpha}$ transforming in $\overline{\mathbf{4}}$.  We take these matter superfields to transform as $(\mathbf{N}, \overline{\mathbf{M}})$ under the gauge group $U(N) \times U(M)$.  The hermitian conjugate fields $\bar{\Phi}_{I} = (\Phi^{I})^{\dagger}$, $\bar{\Psi}^{I} = (\Psi_{I})^{\dagger}$ transform as $(\mathbf{M}, \overline{\mathbf{N}})$.  In terms of algebra generators, the gauge fields have components in the diagonal blocks while the matter fields have components in the off-diagonal blocks.

The derivative $[\nabla_{I J \alpha}, \Phi^{K}]$ transforms in $\mathbf{4} \otimes \mathbf{6} = \overline{\mathbf{4}} \oplus \overline{\mathbf{20}}$.  We impose the constraint that $\overline{\mathbf{20}}$ component vanishes.  In turn, this constrains the form of $W$ as we will show.  We take
\begin{equation}
  [\nabla_{I J \alpha}, \Phi^{K}] = \delta_{[I}^{K} \Psi_{J] \alpha}.
\end{equation}  If we apply to this equation $\nabla_{K L \beta}$ and we use the anticommutation relations derived previously, we find
\begin{equation}
  \label{eq:matter-constr}
  \delta_{[I\vert}^{P} \lbrace \nabla_{K L \beta}, \Psi_{\vert J] \alpha}\rbrace + (I J \alpha) \leftrightarrow (K L \beta) = 2 \epsilon_{I J K L} [\nabla_{\alpha \beta}, \Phi^{P}] + \epsilon_{\alpha \beta} \epsilon_{M I J [K\vert} [W^{M}_{\hphantom{M} \vert L]}, \Phi^{P}].
\end{equation}  If we contract with $\epsilon^{\alpha \beta} \epsilon^{I J K N}$ we find the constraint
\begin{equation}
  \label{eq:W-constraint}
  [W^{(I}_{\hphantom{I} J}, \Phi^{K)}] = \frac 1 5 \delta_{J}^{(I} [W^{K)}_{\hphantom{K} N}, \Phi^{N}].
\end{equation}  This is equivalent to the statement that the
$\mathbf{36}$ in the decomposition $\mathbf{15} \otimes \mathbf{4} =
\mathbf{4} \oplus \mathbf{20} \oplus \mathbf{36}$ of
$[W^{I}_{\hphantom{I} J}, \Phi^{K}]$ is set to zero.

The constraint in eq.~\eqref{eq:matter-constr} relates the fermionic derivative of the spinor superfield to the derivative of the scalar superfield and the commutator $[W, \Phi]$.  In order to find the content of this constraint, we project the LHS and RHS onto the irreps $(\mathbf{4}, \mathbf{1})$, $(\mathbf{4}, \mathbf{3})$, $(\mathbf{20}, \mathbf{1})$ and $(\mathbf{20}, \mathbf{3})$.  Combining these projections with appropriate coefficients we find
\begin{equation}
  \lbrace \nabla_{I J \alpha}, \Psi_{K \beta}\rbrace = -4 \epsilon_{I J K L} \nabla_{\alpha \beta} \Phi^{L} + \frac 1 2 \epsilon_{\alpha \beta} \epsilon_{M I J L} [W^{M}_{\hphantom{M} K}, \Phi^{L}] - \frac 3 {10} \epsilon_{\alpha \beta} \epsilon_{M I J K} [W^{M}_{\hphantom{M} L}, \Phi^{L}].
\end{equation}

Now we want to express $W$ in terms of the fields in our theory.
Given the dimension of $W$, its $R$-symmetry transformations, its
tracelessness, its hermiticity properties and the constraint in
eq.~\eqref{eq:W-constraint}, the answer is heavily constrained.  The
choice which yields the ABJ theory~\cite{Aharony2008a} is
\begin{equation}
  W^{I}_{\hphantom{I} J} = \frac 1 g \left(\Phi^{I} \bar{\Phi}_{J} - \frac 1 4 \delta^{I}_{J} \Phi^{K} \bar{\Phi}_{K} + \bar{\Phi}_{J} \Phi^{I} - \frac 1 4 \delta_{J}^{I} \bar{\Phi}_{K} \Phi^{K}\right).
\end{equation}  The first two terms in $W$ transform as the adjoint of $U(N)$ while the last two transform as the adjoint of $U(M)$.  In order to obtain the ABJM theory we set $N = M$.  Also, the coupling constant $g$ is quantized such that the exponential of the action $e^{i S}$ is gauge invariant.  We will fix the coupling constant later when we discuss the Lagrangian of the theory.

Let us now write the equations of motion of the theory.  The equation of motion for the field strength is obtained from the Bianchi identity of $\nabla_{\alpha \beta}$, $\nabla_{IJ \gamma}$ and $\nabla_{K L \delta}$.  Before we write the equation of motion we should recall that we have two gauge groups: $U(N)$ with gauge field $A$ and $U(M)$ with gauge field $\tilde{A}$.  Therefore, we have two field strengths, $F$ and $\tilde{F}$.  The equations of motion for the gauge fields are:
\begin{align}
  F_{\alpha \gamma} &= \frac 1 {4 g} \left(
  \left[\nabla_{\alpha \gamma },\Phi^I\right] \overline{\Phi }_I-
  \Phi^I\left[\nabla_{\alpha \gamma}, \overline{\Phi}_I\right]-
  \frac{1}{4} \Psi_{I(\alpha} \overline{\Psi}_{\gamma) }^I\right),\\
  \tilde{F}_{\alpha \gamma} &=-\frac 1 {4 g}\left(
  \left[\nabla_{\alpha \gamma},\overline{\Phi }_I\right]\Phi^I-
  \overline{\Phi}_I \left[\nabla_{\alpha \gamma },\Phi ^I\right]-
  \frac{1}{4} \overline{\Psi }_{(\alpha}^I \Psi _{I\gamma)}\right).
\end{align}

The equations of motion for the fermions are obtained from the Bianchi identity for $\nabla_{I J \alpha}$, $\nabla_{K L \beta}$ and $\Psi_{M \gamma}$.  We obtain
\begin{equation}
\epsilon^{\beta \gamma } \left[\nabla_{\alpha \beta },\Psi_{I\gamma }\right] =
\frac{\Phi^J \overline{\Phi}_I \Psi_{J\alpha }}{4 g}-
\frac{\Phi^J \overline{\Phi}_J \Psi_{I\alpha }}{8 g}+
\frac{\Psi_{I\alpha} \overline{\Phi}_J \Phi^J}{8 g}-
\frac{\Psi_{J\alpha} \overline{\Phi}_I \Phi^J}{4 g}+
\frac{\epsilon_{IJKL} \left(\Phi^J \overline{\Psi }_{\alpha }^K \Phi    ^L\right)}{4 g}.
\end{equation}

Finally, the equation of motion for the scalars can be determined from the equation of motion for the fermions, by taking the anticommutator of the fermion equation of motion with $\nabla_{L P \gamma}$, multiplying by $\epsilon^{\alpha \gamma} \epsilon^{L M P Q}$ and using the Jacobi identities and the constraints.  We obtain
\begin{multline}
\epsilon^{\alpha \beta'} \epsilon^{\alpha' \beta } \left[\nabla _{\alpha \alpha' },\left[\nabla _{\beta \beta' },\Phi ^Q\right]\right] = \frac{\epsilon^{\alpha \alpha' } \left(\Phi ^I\medspace \overline{\Psi }_{\alpha }^Q\medspace \Psi _{I\alpha' }\right)}{16 g}-\frac{\epsilon^{\alpha \alpha' } \left(\Phi ^Q\medspace \overline{\Psi }_{\alpha }^I\medspace \Psi _{I\alpha' }\right)}{32 g}+\\
\frac{\epsilon^{\alpha \alpha' } \epsilon^{IJKQ} \left(\Psi _{I\alpha }\medspace \overline{\Phi }_J\medspace \Psi _{K\alpha' }\right)}{16 g}+\frac{\epsilon^{\alpha \alpha' } \left(\Psi _{I\alpha }\medspace \overline{\Psi }_{\alpha' }^I\medspace \Phi ^Q\right)}{32 g}-\frac{\epsilon^{\alpha \alpha' } \left(\Psi _{I\alpha }\medspace \overline{\Psi }_{\alpha' }^Q\medspace \Phi ^I\right)}{16 g}+\\\frac{\Phi ^I\medspace \overline{\Phi }_I\medspace \Phi ^J\medspace \overline{\Phi }_J\medspace \Phi ^Q}{32 g^2}-\frac{\Phi ^I\medspace \overline{\Phi }_I\medspace \Phi ^Q\medspace \overline{\Phi }_J\medspace \Phi ^J}{16 g^2}-\frac{\Phi ^I\medspace \overline{\Phi }_J\medspace \Phi ^J\medspace \overline{\Phi }_I\medspace \Phi ^Q}{16 g^2}+\frac{\Phi ^I\medspace \overline{\Phi }_J\medspace \Phi ^Q\medspace \overline{\Phi }_I\medspace \Phi ^J}{8 g^2}+\\\frac{\Phi ^Q\medspace \overline{\Phi }_I\medspace \Phi ^I\medspace \overline{\Phi }_J\medspace \Phi ^J}{32 g^2}-\frac{\Phi ^Q\medspace \overline{\Phi }_I\medspace \Phi ^J\medspace \overline{\Phi }_J\medspace \Phi ^I}{16 g^2}.
\end{multline}

These equations of motion can be obtained from the Lagrangian\footnote{Actually this Lagrangian is a superfield whose lowest component is the actual Lagrangian.}
\begin{multline}
  \mathcal{L}_{\text{ABJM}} = -\tr ([\nabla_{\mu}, \Phi^{I}] [\nabla^{\mu}, \bar{\Phi}_{I}]) + \frac i 8 \tr (\bar{\Psi}^{\alpha I} [\nabla_{\alpha \beta}, \Psi_{I}^{\beta}]) -\\ 4 g \epsilon^{\mu \nu \lambda} \tr \left(A_{\mu} \partial_{\nu} A_{\lambda} + \frac 2 3 A_{\mu} A_{\nu} A_{\lambda} - \tilde{A}_{\mu} \partial_{\nu} \tilde{A}_{\lambda} - \frac 2 3 \tilde{A}_{\mu} \tilde{A}_{\nu} \tilde{A}_{\lambda}\right) - V_{\text{bos}} + \mathcal{L}_{\text{Yuk}},
\end{multline} where
\begin{align*}
  V_{\text{bos}} = \frac 1 {192 g^{2}} \tr\Bigl(&
\Phi^{I} \overline{\Phi}_{I} \Phi^{J} \overline{\Phi}_{J} \Phi^{K} \overline{\Phi}_{K} +
\Phi^{I} \overline{\Phi}_{J} \Phi^{J} \overline{\Phi}_{K} \Phi^{K} \overline{\Phi}_{I} + \\&4 \Phi^{I} \overline{\Phi}_{J} \Phi^{K} \overline{\Phi}_{I} \Phi^{J} \overline{\Phi}_{K} -
6 \Phi^{I} \overline{\Phi}_{K} \Phi^{J} \overline{\Phi}_{J} \Phi^{K} \overline{\Phi}_{I}\Bigr),\\
  \mathcal{L}_{\text{Yuk}} = -\frac{\epsilon^{\alpha \beta }}{64 g}
  \tr \Bigl(&-\epsilon^{IJKL} \overline{\Phi }_I\medspace \Psi _{J\alpha }\medspace \overline{\Phi }_K\medspace \Psi _{L\beta } +
  \epsilon_{IJKL} \Phi ^I\medspace \overline{\Psi }_{\alpha}^J\medspace \Phi^K\medspace \overline{\Psi }_{\beta}^L+
  \\&2 \overline{\Phi}_I\medspace \Phi ^J\medspace \overline{\Psi}_{\alpha }^I\medspace \Psi_{J\beta}+
  \Phi ^I\medspace \overline{\Phi }_I\medspace \Psi _{J\alpha}\medspace \overline{\Psi }_{\beta}^J-
  2 \Phi^I\medspace \overline{\Phi }_J\medspace \Psi_{I\alpha}\medspace \overline{\Psi }_{\beta}^J-
  \overline{\Phi}_I\medspace \Phi ^I\medspace \overline{\Psi}_{\alpha }^J\medspace \Psi_{J\beta}\Bigr).
\end{align*}

We could rescale the fields \(\Phi \to g^{1/2} \Phi\) and \(\Psi \to
g^{1/2} \Psi\) to pull the coupling constant \(g\) in front of the
Lagrangian.

The gauge part of the action can also be written in spinor language as
\begin{equation}
  \frac 1 2 \epsilon^{\alpha \alpha'} \epsilon^{\beta \delta} \epsilon^{\gamma \beta'} A_{\alpha' \beta'} \left(\partial_{\alpha \beta} A_{\gamma \delta} + \frac 2 3 A_{\alpha \beta} A_{\gamma \delta}\right) = \epsilon^{\mu \nu \lambda} \tr \left(A_{\mu} \partial_{\nu} A_{\lambda} + \frac 2 3 A_{\mu} A_{\nu} A_{\lambda}\right).
\end{equation}  The first form is closer to the language we've been using so far, but the second form is used to find the quantization condition on the coupling.  The quantization of the Chern-Simons coupling is a consequence of the gauge non-invariance of the Lagrangian.  It can be checked that under a gauge transformation $A_{\mu} \to A_{\mu}' = g^{-1} A_{\mu} g + g^{-1} \partial_{\mu} g$ we have
\begin{multline}
  \delta_{\text{gauge}} \left(\epsilon^{\mu \nu \rho} \tr(A_{\mu} \partial_{\nu} A_{\rho} + \frac 2 3 A_{\mu} A_{\mu} A_{\rho})\right) =\\ \partial_{\rho} \left(\epsilon^{\mu \nu \rho} \tr(A_{\mu} \partial_{\nu} g g^{-1})\right) - \frac 1 3 \epsilon^{\mu \nu \rho} \tr(g^{-1} \partial_{\mu} g g^{-1} \partial_{\nu} g g^{-1} \partial_{\rho} g).
\end{multline}  The first term is a total derivative which we will
ignore.  The integral of the second term is quantized (it is an
integer multiple of $8 \pi^{2}$).  In order for the exponential of the
action to be gauge invariant, we need to choose a global coupling of
$\tfrac k {4 \pi}$, where $k$ is an integer (at the same time, we need
to pick an appropriate normalization\footnote{We choose the generators
  of the gauge algebra \(T^{a}\) to be antihermitian.  This choice
  eliminates some factors of \(i\) from the action.} of the gauge generators, which for $SU(N)$ is $\tr(T^{a} T^{b}) = -\tfrac 1 2 \delta^{a b}$).

We should mention that these properties of the Chern-Simons Lagrangian
distinguish it from the $\mathcal{N}=4$ super-Yang-Mills Lagrangian in four
dimensions.  For $\mathcal{N}=4$ super-Yang-Mills, the on-shell Lagrangian can
be written as a descendant of a protected operator (see ref.~\cite{Eden1999}).
Because of the gauge non-invariance, this can not work here.  Presumably this
makes it more challenging to use supersymmetry to simplify perturbative
computations as has been done for example in ref.~\cite{Eden2012b}. In planar
$\mathcal{N}=4$ super-Yang-Mills, polygonal light-like Wilson loops can also be
computed by taking limits of correlation functions of BPS operators, as has been
shown in ref.~\cite{Alday2011}; the analogous analysis for ABJM has been studied
in ref.~\cite{Bianchi2011}.

Now we can work out the components of the gauge connections.  In order
to eliminate the auxiliary fields we use the Harnad \& Shnider gauge
(see ref.~\cite{Harnad1986}) $\theta^{I J \alpha} A_{I J \alpha} = 0$.  Using this gauge condition and the symmetry of the gamma matrices have that $D \equiv \tfrac 1 2 \theta^{I J \alpha} \nabla_{I J \alpha} = \tfrac 1 2 \theta^{I J \alpha} \partial_{I J \alpha}$.  Then, we obtain
\begin{align}
  (1+D) A_{K L \beta} &= 2 \bar{\theta}_{K L}^{\alpha} A_{\alpha \beta} + \epsilon_{\alpha \beta} \bar{\theta}^{\alpha}_{M [K} W^{M}_{\hphantom{M} L]},\\
  D A_{\beta \gamma} &= - \frac 1 4 \theta^{I J}_{(\beta} \lambda_{I J, \gamma)},\\
  D \Phi^{K} &= \frac 1 2 \theta^{K L \alpha} \Psi_{L \alpha},\\
  D \overline{\Phi}_{K} &= \frac 1 2 \bar{\theta}_{K L}^{\alpha} \overline{\Psi}_{\alpha}^{L},\\
  D \Psi_{K \beta} &= -4 \bar{\theta}_{K L}^{\alpha} [\nabla_{\alpha \beta}, \Phi^{L}] - \frac 1 2 \bar{\theta}_{M L \beta} [W^{M}_{\hphantom{M} K}, \Phi^{L}] + \frac 3 {10} \bar{\theta}_{M K \beta} [W^{M}_{\hphantom{M} L}, \Phi^{L}],\\
  D \overline{\Psi}_{\beta}^{K} &= 4 \theta ^{IK\medspace \alpha} \left[\nabla_{\alpha \beta },\overline{\Phi }_I\right] +\frac{1}{2}  \theta_{\beta}^{IJ} \left[\overline{\Phi }_I,W_{\hphantom{K}J}^K\right] +\frac{3}{10} \theta_{\beta}^{IK} \left[\overline{\Phi }_J,W_{\hphantom{J}I}^J\right].
\end{align}

Now we can build the components of the connection in superspace by using the $D$-recursion relations and the initial conditions $A_{\alpha \beta}\vert = a_{\alpha \beta}$, $\Phi^{I}\vert = \phi^{I}$, $\Psi_{I \alpha}\vert = \psi_{I \alpha}$, $A_{I J \alpha}\vert = 0$.  The $\theta$ expansion to the first few orders is
\begin{align}
  \label{eq:expansionF}
  A_{K L \beta} &= \bar{\theta}_{K L}^{\alpha} a_{\alpha \beta} + \frac 1 {8 g} \epsilon_{\beta\alpha} \bar{\theta}^{\alpha}_{K L } \phi^{P} \bar{\phi}_{P} + \frac 1 {2 g} \epsilon_{\beta\alpha} \bar{\theta}^{\alpha}_{M [K } \phi^{M} \bar{\phi}_{L]} + \cdots,\\
  \label{eq:expansionB}
  A_{\beta \gamma} &= a_{\beta \gamma} - \frac 1 {4 g} \left(\theta^{I J}_{(\beta} \psi_{I \vert \gamma)} \bar{\phi}_{J} + \bar{\theta}_{I J (\beta} \phi^{I} \bar{\psi}^{J}_{\gamma)}\right) + \cdots,
\end{align} where we have only written the part of the gauge connection transforming in the adjoint of $U(N)$; the part transforming in the adjoint of $U(M)$ is similar.

Let us list the form of the field strengths
\begin{align}
  \label{eq:fs1}
  F_{I J \alpha, K L \beta} &= \epsilon_{\alpha \beta} \epsilon_{M I J
    [K} W^{M}_{\hphantom{M} L]},\\
  \label{eq:fs2}
  F_{I J \alpha, \beta \gamma} &= \frac 1 3 \left(\epsilon_{\alpha
      \beta} F_{I J}^{\delta}{}_{, \gamma \delta} + \epsilon_{\alpha
      \gamma} F_{I J}^{\delta}{}_{, \beta \delta})\right),\\
  \label{eq:fs3}
  F_{\alpha \beta, \gamma \delta} &= -\frac 1 4 (\epsilon_{\alpha \gamma} F_{\beta \delta} + \epsilon_{\beta \delta} F_{\alpha \gamma} + \epsilon_{\beta \gamma} F_{\alpha \delta} + \epsilon_{\alpha \delta} F_{\beta \gamma}).
\end{align}  It is easy to see that if $\delta x_{\alpha \beta}
= t \lambda_{\alpha} \lambda_{\beta}$ and $\delta \theta^{I J, \alpha}
= \lambda^{\alpha} \eta^{I J}$, then the field strengths vanish when
contracted with $\delta x$ and $\delta \theta$.  What about the
converse statement?  That is, given the flatness conditions on such submanifolds, do they imply the constraints and the equations of
motion?  This is not true.  One missing constraint is the one in
eq.~\eqref{eq:W-constraint}, which arises in the matter sector.
Another missing constraint is the constraint that \(W\) transforms in
\(\mathbf{15}\) of \(SU(4)\).

It would be interesting to look for a formulation where all the
equations of motion arise from flatness conditions on some some
submanifolds of superspace which are well-behaved under superconformal
transformations.  A way to search for such a formulation is to look at
the flag manifolds of \(\mathbb{C}^{4 \vert 6}\) which is a space with
a natural action of the superconformal group \(OSp(6 \vert 4)\) (see
refs.~\cite{MR1632008, Howe:1995md} for the description of the general
theory).  A promising choice seems to be the flag \(\mathbb{C}^{2
  \vert 3} \subset \mathbb{C}^{4 \vert 6}\) which is also
distinguished by the fact that it does not have supersymmetric
torsion, just like chiral superspaces.

Such a formulation, if it exists, would be the basis of a twistorial formulation
of ABJM theory and would probably be a good starting point for studying the dual
of ABJM scattering amplitudes.  The \(\mathcal{N}=4\) super-Yang-Mills
scattering amplitudes enjoy a Yangian symmetry (see ref.~\cite{Drummond:2009fd})
and the same holds for the ABJM scattering amplitudes (see
ref.~\cite{Bargheer:2010hn, Huang2010a}).  However, at strong coupling the fate
of the Yangian symmetry is less clear~\cite{Adam2010a,Bakhmatov2011}.  See also
ref.~\cite{Bianchi2012} for a study of the Wilson-loop/scattering amplitudes
duality in ABJM theory at four points, where the tree-level scattering amplitude
is factored out.

\section{Superwistors for \texorpdfstring{$OSp(6\vert 4)$}{OSp(6 4)}}
\label{sec:twistors-osp64}

In this section we introduce \(OSp(6 \vert 4)\) supertwistors.
We take \(x_{\alpha \beta}\) to be a \(2 \times 2\) matrix and
\(\theta_{\alpha}^{IJ}\) to be a \(2 \times 6\) matrix.  When we write
products of \(x\) and \(\theta\) the matrix product is understood.
The contraction \(\theta \varpi \theta^{T}\) is defined with the help
of the \(SU(4)\) \(\epsilon\) tensor:
\((\theta \varpi \theta^{T})_{\alpha \beta} = \tfrac 1 4
\theta_{\alpha}^{IJ} \epsilon_{IJKL} \theta_{\beta}^{KL}\).  As a
rule, whenever we contract two antisymmetric indices we include a
factor of \(\tfrac 1 2\) to prevent double counting.  For example,
\(\varpi \theta\) is a shorthand notation for \(\tfrac 1 2
\epsilon_{IJKL} \theta^{KL}\).  Notice that the \(2 \times 2\) matrix
\(\theta \varpi \theta^{T}\) is antisymmetric.

We define \(x^{\pm} = x \pm \tfrac 1 2 \theta \varpi \theta^{T}\).  Then,
given \((x^{\pm}, \theta)\) we define two two-planes in \(\mathbb{C}^{4
\vert 6}\) by
\begin{gather}
  \lambda \mapsto (\lambda, \lambda x^{+}, \lambda \theta) \equiv (\lambda,
  \mu, \chi) = Z,\\
  \lambda \mapsto
  \begin{pmatrix}
    -x^{-} \lambda^{T}\\ \lambda^{T} \\ -\varpi \theta^{T} \lambda^{T}
  \end{pmatrix} \equiv
  \begin{pmatrix}
    \tilde{\mu}\\ \lambda^{T}\\ \tilde{\chi}
  \end{pmatrix} = \bar{Z}.
\end{gather}  More properly, the second two-plane lives in the dual
\(\mathbb{C}^{4 \vert 6}\). Then we have
\begin{equation}
  Z \cdot \bar{Z} = \lambda (x^{+} - x^{-} - \theta \varpi \theta^{T})
  \lambda^{T} = 0,
\end{equation} which implies that the two two-planes are orthogonal.
Said differently, the contraction is performed with a symplectic form
which is antisymmetric.  Under a superconformal transformation \(h \in
OSp(6 \vert 4)\) we have \(Z \to Z h\) and \(\bar{Z} \to h^{-1}
\bar{Z}\).  Therefore, simple superconformal invariants can be
obtained by contraction \(Z \cdot \bar{Z}'\).

A choice of \(Z, \bar{Z}\) such that \(Z \cdot \bar{Z} = 0\) yields a
light-like line.  If \(x_{0}^{\pm}, \theta_{0}\) is a particular solution of
the twistor equations such that \(x_{0}^{+}-x_{0}^{-} = \theta_{0} \varpi
\theta_{0}^{T}\), then the general solution is:
\begin{align}
  x^{+} &= x_{0}^{+} + \epsilon \lambda^{T} \eta \varpi \theta_{0}^{T} + t
  \epsilon \lambda^{T} \lambda \epsilon,\\
  x^{-} &= x_{0}^{-} + \theta_{0} \varpi \eta^{T} \lambda \epsilon + t
  \epsilon \lambda^{T} \lambda \epsilon,\\
  \theta &= \theta_{0} + \epsilon \lambda^{T} \eta, \qquad
  \theta^{T} = \theta_{0}^{T} - \eta^{T} \lambda \epsilon,\\
  x &= x_{0} + t \epsilon \lambda^{T} \lambda \epsilon + \frac 1 2
  (\epsilon \lambda^{T} \eta \varpi \theta_{0}^{T} + \theta_{0} \varpi
  \eta^{T} \lambda \epsilon),
\end{align} where \(\epsilon\) is a \(2 \times 2\) antisymmetric
matrix such that \(\epsilon^{2} = -1\). Therefore, we see that a
light-like line has dimension \((1 \vert 6)\).  The translations along
this ``fat'' line are generated by the following vector fields
\begin{gather}
  \epsilon \lambda^{T} \lambda \epsilon \frac \partial {\partial x},\\
  \epsilon \lambda^{T} \eta \frac \partial {\partial \theta} + \frac 1
  2 \epsilon \lambda^{T} \eta \varpi \theta^{T} \frac \partial {\partial x} +
  \frac 1 2 \theta \varpi \eta^{T} \lambda \epsilon \frac \partial {\partial
    x} = \epsilon \lambda^{T} \eta \left(\frac \partial {\partial
      \theta} + \varpi \theta^{T} \frac \partial {\partial x}\right),
\end{gather} where we have used the fact that \(x\) is symmetric as a
matrix.  Above, we recognize the SUSY covariant derivative
\(D\).

Let us introduce the bosonic and fermionic vielbeine.  The total
derivative in variables \((x, \theta)\) is given by
\begin{equation}
  \label{eq:tot-der}
  d = d \theta \cdot \frac {\partial}{\partial \theta} + d x \cdot
  \frac {\partial}{\partial x},
\end{equation} where \(\cdot\) stands for total contraction of
indices.  This total derivative can also be written as
\begin{equation}
  d = e_{B} \cdot \frac {\partial}{\partial x} + e_{F} \cdot D,
\end{equation} where \(D\) is the fermionic SUSY covariant
derivative.  Identifying terms we obtain
\begin{equation}
  e_{B} = d x - \frac 1 2 d \theta \varpi \theta^{T} + \frac 1 2
  \theta \varpi d \theta^{T}, \qquad e_{F} = d \theta,
\end{equation} where we have symmetrized \(e_{B}\) since it is
contracted with a symmetric tensor.

Two light-like lines given by \((Z, \bar{Z})\) and \((Z', \bar{Z}')\)
respectively, intersect if \(Z \cdot \bar{Z}' = 0\) and \(Z' \cdot
\bar{Z} = 0\).  We also have \(Z \cdot \bar{Z} = 0\) and \(Z' \cdot
\bar{Z}' = 0\), as explained above.

The considerations above also lead to a natural definition of
supersymmetry invariant distance between points \((x_{i},\theta_{i})\) and
\((x_{j}, \theta_{j})\)
\begin{equation}
  \label{eq:susy-distance}
  \Delta x_{ij} = x_{i}^{+} - x_{j}^{-} - \theta_{i} \varpi \theta_{j}^{T},
\end{equation} which is such that \(Z_{i} \cdot \bar{Z}_{j} =
\lambda_{i} \Delta x_{ij} \lambda_{j}^{T}\).  Taking into account that
\((\Delta x_{ij})^{T} = -\Delta x_{ji}\), we find that \(Z_{i} \cdot
\bar{Z}_{j} = - Z_{j} \cdot \bar{Z}_{i}\).  There are other ways to
define intervals which are invariant under super-Poincar\'e
transformations.  They all differ by nilpotent terms.  The \(\Delta
x_{ij}\) defined above is the three-dimensional counterpart of a
chiral--antichiral interval.  It has the peculiarity that
\(\Delta x_{ij} \neq -\Delta x_{ji}\).

We can define another super-Poincar\'e invariant interval \(\delta
x_{ij}\) which is such that \(\delta x_{ij} = -\delta x_{ji}\) by taking
\begin{equation}
  \delta x_{ij} = \frac 1 2 (\Delta x_{ij} - \Delta x_{ji}) = x_{i} -
  x_{j} - \frac 1 2 \theta_{i} \varpi \theta_{j}^{T} + \frac 1 2
  \theta_{j} \varpi \theta_{i}^{T}.
\end{equation}  We also have \(\delta x_{ij}^{T} = \delta x_{ij}\).

The configuration of pairwise light-like separated points in superspace can be
given as a sequence of twistors \(Z_{i}, \bar{Z}_{i}\) such that \(Z_{i} \cdot
\bar{Z}_{i} = 0\), \(Z_{i} \cdot \bar{Z}_{i \pm 1}= 0\).  We solve for \(x_{i},
\theta_{i}\) from the components of \(Z_{i}\) and \(\bar{Z}_{i}\).  Using
\begin{gather}
  \mu_{i} = \lambda_{i} x_{i}^{+}, \qquad
  \mu_{i+1} = \lambda_{i+1} x_{i}^{+},\\
  \tilde{\mu}_{i} = -x_{i}^{-} \lambda_{i}^{T}, \qquad
  \tilde{\mu}_{i+1} = -x_{i}^{-} \lambda_{i+1}^{T},\\
  \chi_{i} = \lambda_{i} \theta_{i}, \qquad
  \chi_{i+1} = \lambda_{i+1} \theta_{i},
\end{gather} we find
\begin{gather}
  x_{i}^{+} = \frac {\epsilon \lambda_{i+1}^{T} \mu_{i} - \epsilon
    \lambda_{i}^{T} \mu_{i+1}}{\lambda_{i} \epsilon
    \lambda_{i+1}^{T}},\\
  x_{i}^{-} = \frac {\tilde{\mu}_{i} \lambda_{i+1} \epsilon -
    \tilde{\mu}_{i+1} \lambda_{i} \epsilon}{\lambda_{i} \epsilon
    \lambda_{i+1}^{T}},\\
  \theta_{i} = \frac {\epsilon \lambda_{i+1}^{T} \chi_{i}-\epsilon
    \lambda_{i}^{T} \chi_{i+1}}{\lambda_{i} \epsilon
    \lambda_{i+1}^{T}}.
\end{gather}  Geometrically, these equations describe the point
\((x_{i}, \theta_{i})\) as the intersection of two light-like lines
represented by \((Z_{i}, \bar{Z}_{i})\) and \((Z_{i+1},
\bar{Z}_{i+1})\), respectively.

It is easy to show that
\begin{gather}
  \theta_{i} - \theta_{i+1} = \epsilon \lambda_{i+1}^{T} \eta_{i+1},
\end{gather} where
\begin{gather}
  \eta_{i} = \frac {\langle i-1 i\rangle \chi_{i+1} + \langle i
    i+1\rangle \chi_{i-1} + \langle i+1 i-1\rangle \chi_{i}}{\langle
    i-1 i\rangle \langle i i+1\rangle}
\end{gather} and \(\langle i j\rangle = \lambda_{i} \epsilon
\lambda_{j}^{T}\).

In order to compute the distance squared \((\Delta x_{ij})^{2}\) we
first compute the matrix elements
\begin{gather}
  \lambda_{i} \Delta x_{ij} \lambda_{j}^{T} = Z_{i} \cdot \bar{Z}_{j},
  \qquad
  \lambda_{i} \Delta x_{ij} \lambda_{j+1}^{T} = Z_{i} \cdot
  \bar{Z}_{j+1},\\
  \lambda_{i+1} \Delta x_{ij} \lambda_{j}^{T} = Z_{i+1} \cdot \bar{Z}_{j},
  \qquad
  \lambda_{i+1} \Delta x_{ij} \lambda_{j+1}^{T} = Z_{i+1} \cdot
  \bar{Z}_{j+1}.
\end{gather}  From this we find that
\begin{equation}
  - (\Delta x_{ij})^{2} = \det \Delta x_{ij} = \frac {(Z_{i} \cdot \bar{Z}_{j}) (Z_{i+1} \cdot
    \bar{Z}_{j+1}) - (Z_{i} \cdot \bar{Z}_{j+1}) (Z_{i+1} \cdot
    \bar{Z}_{j})}{\langle i i+1\rangle \langle j j+1\rangle}.
\end{equation}

Above we mentioned that \(\Delta x_{ij} \neq -\Delta x_{ji}\).
However, for a light-like interval \(\Delta x_{i,i+1} = \Delta
x_{i+1,i}\) due to \((\theta_{i}-\theta_{i+1}) \varpi
(\theta_{i}-\theta_{i+1})^{T} = 0\).  As a consequence \(\Delta
x_{i,i+1} = \delta x_{i,i+1}\).

Not all the quantities we will encounter can be expressed only in terms of
twistors since they are not conformal invariant.  For writing down
quantities which are Lorentz but not conformal invariant we introduce
an ``infinity'' twistor \(\mathcal{I}\), which is such that \(Z_{i}
\mathcal{I} \bar{Z}_{j} = \langle i j\rangle\), \(\str (X_{i}
\mathcal{I}) = -2 \langle i i+1\rangle\).  In terms of \((4\vert
6) \times (4\vert 6)\) matrices the infinity twistor is
\begin{equation}
  \label{eq:inf-twistor}
  \mathcal{I} = \begin{pmatrix}
    0 & \epsilon & 0\\
    0 & 0 & 0\\
    0 & 0 & 0
  \end{pmatrix}.
\end{equation}  This infinity twistor is also preserved (that is
commutes with) the Poincar\'e supersymmetry.  Conformal inversion acts
as \(X \to \Upsilon X \Upsilon\), where
\begin{equation}
  \label{eq:inv-matrix}
  \Upsilon = \begin{pmatrix}
    0 & 1 & 0\\
    1 & 0 & 0\\
    0 & 0 & 1
  \end{pmatrix}.
\end{equation}  The zero twistor, which is obtained by setting
\(x=0\), \(\theta=0\) is obtained from the infinity twistor
\(\mathcal{I}\) by inversion.

We are also interested in parametrizing the light-like segment between
points with coordinates \((x_{i}, \theta_{i})\) and \((x_{i+1},
\theta_{i+1})\).  A point represented by the two pairs of twistors
\((Z_{i+1}, \bar{Z}_{i+1})\) and \((\alpha Z_{i}+\beta Z_{i+2}, \alpha
\bar{Z}_{i}+\beta \bar{Z}_{i+2})\) with \(\alpha = (1-\tau) \langle
i+2,i+1\rangle\), \(\beta = \tau \langle i i+1\rangle\) lies on the
segment \((i,i+1)\).  First of all, notice that these two pairs of
twistors parametrize intersecting light-like lines since \(Z_{i+1}
\cdot (\alpha Z_{i} + \beta Z_{i+2}) = 0\).  Then, solving for
\(x^{\pm}, \theta\) we find \(x^{\pm} = (1-\tau) x_{i}^{\pm} + \tau
x_{i+1}^{\pm}\), \(\theta = (1-\tau) \theta_{i} + \tau \theta_{i+1}\)
which also implies that \(x = (1-\tau) x_{i} + \tau x_{i+1}\).  Then
we compute the vielbeine along this curve and we obtain\footnote{The
  formula for \(e_{B}\) implicitly contains a choice of normalization
  for \(\lambda_{i+1}\).  The expression which is invariant under
  rescaling is given by \(e_{B} = \tfrac {(\mathcal{I} \bar{Z}_{i+1})
    (Z_{i+1} \mathcal{I})}{\langle i i+1 \mathcal{I}\rangle \langle
    i+1 i+2 \mathcal{I}\rangle} (Z_{i} \cdot \bar{Z}_{i+2})\), where
  we have written all the infinity twistors explicitly \(\langle i i+1
  \mathcal{I}\rangle = \langle i i+1\rangle\) to show that \(e_{B}\)
  is invariant also under the rescaling of the infinity twistor.  In
  the following we will assume that the \(\lambda_{i}\) have been
  normalized such that the bosonic vielbein on side \(i\) is given by
  \(e_{B} = \epsilon \lambda_{i+1}^{T} \lambda_{i+1} \epsilon\).}
\begin{gather}
  e_{B} = d \tau (x_{i+1} - x_{i} - \frac 1 2 \theta_{i+1} \varpi
  \theta_{i}^{T} + \frac 1 2 \theta_{i} \varpi \theta_{i+1}^{T}) = d
  \tau \delta x_{i+1,i} = d \tau \epsilon \lambda_{i+1}^{T}
  \lambda_{i+1} \epsilon, \\
  e_{F} = d \tau (\theta_{i+1} - \theta_{i}) = d \tau \epsilon
  \lambda_{i+1}^{T} \eta_{i+1}.
\end{gather}

There is an obvious ambiguity in parametrizing a point \(x_{i}\) as the
intersection of two light-like lines since one can choose some other
light-like lines with the same intersection.  Instead of using the
lines determined by \((Z_{i}, \bar{Z}_{i})\) and \((Z_{i+1},
\bar{Z}_{i+1})\), we can choose the lines determined by \((Z_{i}',
\bar{Z}_{i}')\) and \((Z_{i+1}', \bar{Z}_{i+1}')\), where the primed
quantities are determined from the unprimed ones by a \(2 \times 2\)
matrix.  However, the quantity\footnote{Writing the indices explicitly
this becomes \({X_{i}}_{A}^{\hphantom{A} B} = \bar{Z}_{i,A}
Z_{i+1}^{B} - \bar{Z}_{i+1,A} Z_{i}^{B}\).  Note that the fermionic
components of \(Z_{i}\) and \(Z_{i+1}\) do not commute and the order
is important.}
\begin{equation}
  X_{i} = \bar{Z}_{i} \otimes Z_{i+1} - \bar{Z}_{i+1} \otimes Z_{i}
\end{equation} remains unchanged up to a rescaling by the determinant
of the \(2 \times 2\) transformation matrix.  In order to make
quantities which are independent on the scaling we have to make sure
that these determinant factors cancel.  For example, the ratio
\(X_{i}/\langle i i+1\rangle\) is invariant under rescaling.

Using the fact that \(\lambda_{i}^{T} \otimes \lambda_{i+1} -
\lambda_{i+1}^{T} \otimes \lambda_{i} = \epsilon \langle i
i+1\rangle\), which can be obtained by considering contractions with
\(\epsilon \lambda_{i}^{T}\) and \(\epsilon \lambda_{i+1}^{T}\) at the
right, we find
\begin{equation}
  \label{eq:X-matrix}
  {X_{i}}_{A}^{\hphantom{A} B} = \langle i i+1\rangle
  \begin{pmatrix}
    -x_{i}^{-} \epsilon & -x_{i}^{-} \epsilon x_{i}^{+} & -x_{i}^{-}
    \epsilon \theta_{i}\\
    \epsilon & \epsilon x_{i}^{+} & \epsilon \theta_{i}\\
    -\varpi \theta_{i}^{T} \epsilon & -\varpi \theta_{i}^{T} \epsilon x_{i}^{+} &
    -\varpi \theta_{i}^{T} \epsilon \theta_{i}
  \end{pmatrix}.
\end{equation}  The \((4 \vert 6) \times (4 \vert 6)\) matrix
\(X_{i}\) is such that \((X_{i})^{2} = 0\) and \(\str X_{i} = 0\).

From the definition it is clear that under superconformal
transformations \(h \in OSp(6 \vert 4)\) a matrix \(X\) transforms as
\(X \to h X h^{-1}\).  Then, given two points \(X_{i}\), \(X_{j}\), we
can construct an invariant \(\str X_{i} X_{j}\).  Explicit computation
yields
\begin{equation}
  \str X_{i} X_{j} = 2 \langle i i+1\rangle \langle j j+1\rangle \det
  \Delta x_{ij} = -2 \langle i i+1\rangle \langle j j+1\rangle (\Delta
  x_{ij})^{2}.
\end{equation}

A three-point invariant is given by
\begin{equation}
  \frac {\str (X_{i} X_{j} X_{k})}{\langle i i+1\rangle \langle j
    j+1\rangle \langle k k+1\rangle} = \tr (\epsilon \Delta x_{ij}
  \epsilon \Delta x_{jk} \epsilon \Delta x_{ki}),
\end{equation} which is a supersymmetrization of \(\epsilon_{\mu \nu
  \rho} x_{ij}^{\mu} x_{jk}^{\nu} x_{ki}^{\rho} \sim \epsilon_{\mu \nu
\rho} x_{i}^{\mu} x_{j}^{\nu} x_{k}^{\rho}\).

So far have taken \(Z\) and \(\bar{Z}\) to have complex components.
However, the \(OSp(6 \vert 4)\) elements are represented by \((4 \vert
6) \times (4 \vert 6)\) supermatrices with real entries.  We want the
matrix \(X_{i}/\langle i i+1\rangle\) to have real entries.  This does
not mean that the components of the twistors \(Z\) should be real.  In
fact, some of them have to be imaginary in order for the Wilson loop
to close.

Later we will compute perturbative corrections to a polygonal
light-like Wilson loop in superspace.  This is a sum of diagrams with
gluon and matter exchanges between the sides of the polygon and
interaction vertices which are integrated over.  There are several
ways to parametrize the sides.  We can take
\begin{equation}
  X(\tau) = (1-\tau) \frac {X_{i}}{\langle i i+1\rangle} + \tau \frac {X_{i+1}}{\langle i+1 i+2\rangle},
\end{equation} where \(\tau \in [0,1]\).  This parametrization is
arranged such that \(\str (X(\tau) \mathcal{I}) = 1\).  Then, the parts
in the denominator of the propagator which involve the infinity
twistor cancel out, but the rest of the terms become more complicated
(and still depend on the infinity twistor). Another parametrization we
can choose is\footnote{If we insist on using kinematics where the
  coordinates of the Wilson loop vertices are real, then some of the
  \(\lambda_{i}\) have to be imaginary.  So in some cases \(\tau\)
  goes from zero to infinity along the imaginary axis.}
\begin{equation}
  X(t) = X_{i} + t X_{i+1},
\end{equation} with \(t \in [0,\infty)\).

Some of the diagrams contain integration over the space-time.  Even
though we will not consider in detail such diagrams in this paper, it
is useful to work out the twistor representation of the integration
measure.  The superconformal integration measure in variables \((x,
\theta)\) can be obtained as follows.  To the point in superspace
with coordinates \((x, \theta)\) we can associate two
supertwistors \((Z_{A}, Z_{B})\) (here \(A\) and \(B\) are labels, not
components of the a twistor \(Z\)).  The over-lined versions of these
supertwistors are not independent but can be obtained by transposition
followed by multiplication by a constant matrix.  These supertwistors
satisfy the constraints \(Z_{A} \cdot \bar{Z}_{A} = 0\), \(Z_{A} \cdot
\bar{Z}_{B} = 0\), \(Z_{B} \cdot \bar{Z}_{A} = 0\) and \(Z_{B}
\cdot \bar{Z}_{B} = 0\).  The only independent nontrivial constraint
is \(Z_{A} \cdot \bar{Z}_{B}=0\).  The choice of \((Z_{A},Z_{B})\)
is not unique; any other choice obtained by a \(GL(2)\)
transformation yields the same point in superspace.  Moreover, the
\(GL(2)\) transformations preserve the constraint.
Therefore, we need to divide by the action of this group.  Using these
ingredients we get
\begin{equation}
  d^{3 \vert 12} X = \frac {d^{4 \vert 6} Z_{A} d^{4 \vert 6}
    Z_{B}}{\text{vol}(GL(2))} \delta (Z_{A} \cdot
  \bar{Z}_{B}).
\end{equation}

More precisely, the measure \(d^{4 \vert 6} Z_{A} d^{4 \vert 6} Z_{B}
\delta (Z_{A} \cdot \bar{Z}_{B})\) is invariant under \(SL(2)\) but
not under the \(GL(2)\).  In order to study the invariance under a
global rescaling \(X \to \lambda X\) we can rescale \(Z_{A}
\to \lambda Z_{A}\) and leave the \(Z_{B}\) invariant.  Then, \(d^{4
  \vert 6} Z_{A} \to \lambda^{-2} d^{4 \vert 6} Z_{A}\) and a further
contribution from \(\delta\) produces a factor of \(\lambda^{-3}\).
Then we have that \(d^{3 \vert 12} X \to \lambda^{-3} d^{3 \vert 12}
X\).  Without supersymmetry we would have a conformal invariant
measure such that \(d^{3} X \to \lambda^{3} d^{3} X\).  When gauging
the extra degree of freedom in \(GL(2)/SL(2)\) we consider that
the measure is multiplied by a function such that this extra charge
vanishes.  We should further note that for \(\mathcal{N}=3\)
supersymmetries the measure is exactly invariant under \(GL(2)\) transformations.

This superconformal measure can be written in terms of
\((x,\theta)\) variables by gauging \(\lambda_{A} = (1,0)\)
and \(\lambda_{B} = (0,1)\).  If we restrict to the bosonic case for
simplicity and we set \(Z_{A} = (1,0,\mu_{A})\) and \(Z_{B} =
(0,1,\mu_{B})\) then we can remove the group \(GL(2)\) and we are
left with
\begin{equation}
  d^{2} \mu_{A} d^{2} \mu_{B} \delta(\mu_{A,2} - \mu_{B,1}).
\end{equation}  In this gauge the \(\mu\) components are given by
\(\mu_{A} = (x_{11}, x_{12})\) and \(\mu_{B} = (x_{21}, x_{22})\)
and the constraint imposes the symmetry of the matrix \(x\).  Therefore,
\begin{equation}
  d^{3} x = \frac {d^{4} z_{A} d^{4} z_{B}}{\text{vol}(GL(2))} \delta (z_{A} \cdot \bar{z}_{B}),
\end{equation} where we have denoted by lowercase letters the bosonic
components of the supertwistors \(Z\).  We have also computed the
normalization factor of the twistorial measure.

\section{Super-Wilson loops}
\label{sec:super-wilson-loops}

Let us now define the super-Wilson loops and show that they are
classically invariant under superconformal transformations.

In order to define the polygonal super-Wilson loop we need a choice of
contour and a connection.  The superspace connection is given by
\(\mathcal{A} = e_{B} \cdot A_{B} + e_{F} \cdot A_{F}\), where
\(e_{B}\), \(e_{F}\) are the bosonic and the fermionic vielbeine and
\(A_{B}\), \(A_{F}\) are the bosonic and fermionic connection.  This
connection \(\mathcal{A}\) transforms nicely under super-gauge
transformations.

The contour can be described by giving a sequence of light-like lines
which intersect pairwise.  As we described in
sec.~\ref{sec:twistors-osp64}, each light-like line is parametrized by
a pair of twistors \((Z_{i}, \bar{Z}_{i}) \) such that \(Z_{i} \cdot
\bar{Z}_{i} = 0\).  The conditions that the neighboring sides \(i\) and
\(i+1\) of the polygon intersect is encoded in the constraint \(Z_{i}
\cdot \bar{Z}_{i+1} = 0\) which also implies \(Z_{i+1} \cdot
\bar{Z}_{i} = 0\).

However, as we also discussed in sec.~\ref{sec:twistors-osp64}, in
space-time the right notion of light-like line has several fermionic
directions (it is ``fat'' in the language of refs.~\cite{Beisert2012a,
  Beisert2012}).  These fat lines intersect in points in superspace,
which are the vertices of the polygonal contour.

Since the superspace connection \(\mathcal{A}\) is a \(1\)-form, it
needs to be integrated over a one-dimensional curve, but which one?
The answer, which was given in ref.~\cite{Beisert2012a}, is that, as
long as the contour lies in the ``fat'' lines, it does not matter
which contour we choose since the gauge field is flat there.  In fact,
for any choice of contour a conformal transformation will not preserve
it (it only preserves the ``fat'' line) so in order to show invariance
under conformal transformations we have to deform the contour to the
previous one, while staying inside the ``fat'' line.

In the case of ABJM theory the field strength is flat on the ``fat''
lines, as one can see by examining the field strengths in
eqs.~\eqref{eq:fs1}~\eqref{eq:fs2}~\eqref{eq:fs3}.

Given the polygonal contour \(\mathcal{C}\) we can define the
super-Wilson loop as
\begin{equation}
  \mathcal{W} = \left\langle \tr P \exp \left(\int_{\mathcal{C}}
    \mathcal{A}\right)\right\rangle.
\end{equation}  Recall that the gauge connection \(\mathcal{A}\) lives
in the \((\mathbf{Ad}, \mathbf{1})\oplus (\mathbf{1},\mathbf{Ad})\)
of \(U(N) \times U(M)\).  Then,
the trace above is taken in the \(\mathbf{N} \oplus \mathbf{M}\)
representation.  However, this is not the only option; one could
instead consider the supertrace (see the discussions in
refs.~\cite{Drukker2008, Chen2010, Rey2009, Drukker2010, Cardinali2012, Kim2013, Bianchi2014}).

We will restrict to the planar limit, where \(N \to \infty\), \(M \to
\infty\), \(k \to \infty\) such that the ratios \(\tfrac N k\) and
\(\tfrac M k\) are constant.  From our definition of the super-Wilson
loop it follows that the \(U(N)\) and the \(U(M)\) parts do not mix.
More explicitly, at the leading order in \(N\) and \(M\) all the
Feynman graphs attach either to the \(U(N)\) or to the \(U(M)\) part
of the super-Wilson loop.  Since there are always two possibilities
for the color factors and since they are very simple, we will not
write them out explicitly.

Notice that the \(\theta\) expansion of the connections in
eqs.~\eqref{eq:expansionF}~\eqref{eq:expansionB} contains composite
fields, i.e.\ products of local fields at the same space-time point.
Such products are singular in the quantum theory and need to be
normal-ordered.  In fact, each light-like side needs to be
normal-ordered since any contraction between fields on each side is
singular.  We will not attempt to give a prescription for how to do
this.

The fact that at leading order in the color factors the \(U(N)\) and
\(U(M)\) factors do not mix is not a satisfactory feature.  One can
introduce mixing in several ways.  For example, we can insert
bifundamental fields at the vertices.  Another way, inspired by the
construction (see ref.~\cite{Drukker2010}) of the \(\tfrac 1 2\)-BPS
Wilson loop in ABJM theory, is to use a super-connection\footnote{This
idea arose in conversations with Niklas Beisert.} in the sense
that the gauge part is a \((N\vert M) \times (N\vert M)\) supermatrix.

A natural question is whether the supersymmetric Wilson loop we construct in
this paper is dual to scattering amplitudes.  The answer is negative for several
reasons.  First, the symmetries do not match; the scattering amplitudes only
preserve an \(SU(3) \times U(1)\) out of the \(SO(6)\) R-symmetry, while the
supersymmetric Wilson loop we construct preserves the full \(SO(6)\) symmetry.
Second, the supersymmetric Wilson loop can be constructed for any number of
sides while scattering amplitudes only exist for an even number of external
particles.  This can be cured by introducing operators transforming in
bi-fundamental representations at the vertices of the Wilson loop.  These
operators can also naturally break the symmetry \(SO(6) \to SU(3) \times U(1)\).
The third reason why a duality with scattering amplitudes is not possible for
the supersymmetric Wilson loops we study in this paper is the non-chiral nature
of the \(\mathcal{N}=6\) superspace.  The analog problem for \(\mathcal{N}=4\)
super-Yang-Mills was studied in ref.~\cite{Beisert2012a, Beisert2012}.  There it
was shown that the superspace torsion prevents an identification between the
kinematics of supersymmetric Wilson loops and scattering amplitudes, since it is
not possible to define supersymmetric intervals that are both null and sum to
zero.  Chiral superspace is torsionless and there this obstruction disappears.

\section{Some perturbative computations}
\label{sec:pert-comp}

What is the $\theta$ dependence of the supersymmetric Wilson loop?  At
the lowest order we need to compute a correlation between the terms
inside $A_{\alpha \beta}$ which are linear in $\theta$.  We therefore
need to compute correlation functions of type $\langle \psi(x)
\bar{\phi}(x) \phi(y) \bar{\psi}(y)\rangle = \langle \psi(x) \bar{\psi}(y)\rangle \langle \bar{\phi}(x) \phi(y)\rangle$.

If we drop the color dependence the two-point functions for the scalars and fermions are
\begin{align}
  \langle \bar{\phi}_{I}(x) \phi^{J}(y)\rangle &= -\frac 1 {4 \pi} \delta_{I}^{J} \frac 1 {\lvert x-y\rvert},\\
  \langle \psi_{J}^{\alpha}(x) \bar{\psi}^{I \beta}(y)\rangle &= -\frac 1 {8 \pi} \delta_{J}^{I} \frac {(x-y)^{\alpha \beta}}{\lvert x-y\rvert^{3}}.
\end{align}  Due to the normalization $\tr T^{a} T^{b} = -\tfrac 1 2 \delta^{a b}$ of the gauge algebra generators, the color contribution can be obtained by multiplying by $-2$ for each propagator and by $N$ for every loop transforming in the fundamental of $U(N)$ and by $M$ for every loop transforming in the fundamental of $U(M)$.

We have several types of contributions: between the bilinear transforming in
the adjoint of \(U(N)\) and itself, subleading contributions between
the bilinears transforming in the adjoint of \(U(N)\) with \(U(M)\),
etc.  Only the color structures are different between the leading color
contributions.  We will focus on the dependence on the odd variables.

\begin{figure}
  \centering
  \includegraphics{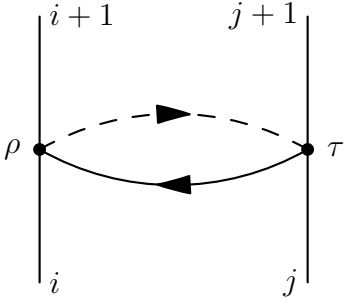}
  \caption{The lowest order contribution in the coupling to the
    \(\chi_{i+1} \cdot \tilde{\chi}_{j+1}\) coefficient of the
    super-Wilson loop.  The fermion propagators are solid lines while
    scalar propagators are dashed lines.}
  \label{fig:ff}
\end{figure}

When restricted to the light-like segment between points labeled by
\(i\) and \(i+1\) the terms in the superconnection which are linear in
\(\theta\) contribute (see fig.~\ref{fig:ff})
\begin{multline}
  -\frac i {4 g} d \tau \lambda_{i+1}^{\beta} \lambda_{i+1}^{\gamma}
  \left(\theta_{\beta}^{IJ}(\tau) \psi_{I \gamma}(\tau) \bar{\phi}_{J}(\tau)
    + \bar{\theta}_{IJ \beta}(\tau) \phi^{I}(\tau) \psi_{\gamma}^{J}(\tau)\right)
  =\\= -\frac {i d \tau}{4 g} \left(\chi_{i+1}^{IJ}
    \lambda_{i+1}^{\gamma} \psi_{I \gamma}(\tau) \bar{\phi}_{J}(\tau) +
    \tilde{\chi}_{i+1,IJ} \phi^{I}(\tau) \bar{\psi}_{\gamma}(\tau)
    \lambda_{i+1}^{\gamma}\right),
\end{multline} where \(\theta_{i}(\tau) = (1-\tau) \theta_{i} + \tau
\theta_{i+1}\) and \(\phi(\tau) = \phi((1-\tau) x_{i} + \tau
x_{i+1})\), etc.

If we compute the correlation between the parts linear in \(\theta\)
on the sides \((i,i+1)\) and \((j,j+1)\), where the side \((j,j+1)\)
is parametrized by \(\sigma\), we find
\begin{equation}
  -\frac 1 {(16 \pi g)^{2}} \frac {(\chi_{i+1} \cdot
    \tilde{\chi}_{j+1}) (z_{i+1} \cdot \tilde{z}_{j+1})}{(((1-\tau) x_{i} +
    \tau x_{i+1} - (1-\sigma) x_{j} - \sigma x_{j+1})^{2})^{2}},
\end{equation} where we have denoted by \(z_{i}\) the bosonic
components of the supertwistor \(Z_{i}\).

We further need to integrate this over \(\tau\) and \(\sigma\) from
\(0\) to \(1\).  The integral is easy to do, and we finally
get\footnote{This result holds if \(i\) and \(j\) are not neighbors,
  i.e.\ \(i \neq j, j \pm 1\).  If they are neighbors, then the
  integrand is exactly zero.  This is unlike in the case of the
  \(\mathcal{N}=4\) super-Yang-Mills super-Wilson loop, where such
  contributions are UV-divergent.}
\begin{equation}
  \frac 1 {(16 \pi g)^{2}} \frac {(\chi_{i+1} \cdot
    \tilde{\chi}_{j+1})}{z_{i+1} \cdot \tilde{z}_{j+1}} \ln \left(\frac
    {x_{i,j+1}^{2} x_{i+1,j}^{2}}{x_{ij}^{2} x_{i+1,j+1}^{2}}\right),
\end{equation} where we have used the identity \((z_{i+1} \cdot
\tilde{z}_{j+1})^{2} = x_{ij}^{2} x_{i+1,j+1}^{2} - x_{i,j+1}^{2} x_{i+1,j}^{2}\).

This answer is a component of the expansion of the \(n\)-sided
light-like supersymmetric Wilson loop \(\mathcal{W}(Z_{1}, \dotsc,
Z_{n})\).  Superconformal invariance dictates that \(\mathcal{W}\)
should depend\footnote{The terms of type \(z_{i} \cdot \tilde{z}_{j}\)
are not all independent.  They are related by Pfaffian constraints,
which are the analog of Gram determinant constraints for an
antisymmetric ``metric''.  As an example of constraint, consider six
bosonic twistors \(z_{i}\), \(i=1,\dotsc,6\) and form the
antisymmetric matrix \(\mathcal{M} = (z_{i} \cdot
\tilde{z}_{j})_{i,j=1,\dotsc,6}\).  Then, \(\pf \mathcal{M} = 0\)
which imposes cubic constraints on the products \(z_{i} \cdot
\tilde{z}_{j}\).} only on terms like \(Z_{i} \cdot \bar{Z}_{j}\) and terms
of type \(\chi_{i} \cdot \tilde{\chi}_{j}\) can only originate from
\(Z_{i} \cdot \bar{Z}_{j}\) terms.  Therefore, the coefficient of
\(\chi_{i} \cdot \tilde{\chi}_{j}\) can be written as
\begin{equation}
  \text{Coefficient of \(\chi_{i} \cdot \tilde{\chi}_{j}\) in
    \(\mathcal{W}\)} = \left.\frac {\partial \mathcal{W}}{\partial Z_{i}
    \cdot \bar{Z}_{j}}\right|_{\chi = 0}.
\end{equation}  Using the result computed above we find that
\begin{equation}
  \left. d \mathcal{W}\right|_{\chi = 0} \propto \frac 1 {(16 \pi
    g)^{2}} \ln \left(\frac {x_{i,j+1}^{2} x_{i+1,j}^{2}}{x_{ij}^{2}
 x_{i+1,j+1}^{2}}\right) d \ln (z_{i+1} \cdot \tilde{z}_{j+1}).
\end{equation}  The proportionality factor is there because we have not
included the color factors.  This result can be thought as a
differential equation which can be integrated in terms of
dilogarithms.\footnote{Of course, this method will not reproduce an
  additive constant.  The constant can be obtained by using the
  properties of the Wilson loop under different degenerations.}

Up to coupling factors this result is exactly the same\footnote{This
  form can be found for example in
  ref.~\cite[eqs.~6.12,6.13]{Beisert2012}} as for \(1\)-loop Wilson
loop in \(\mathcal{N}=4\) super-Yang-Mills, but here it appears at two
loops.  The first computation of this answer in ABJM theory was done
in ref.~\cite{Wiegandt2011}.  Our derivation of the same result is
simpler.

We see that we have a mixing between contributions of different
complexities but whose dependence on the coupling constant is the
same.  For example, two-loop diagrams with gauge fields and diagrams
with matter fields which are of the same complexity as one loop,
contribute at the same order in perturbation theory and are related by
supersymmetry.

\begin{figure}
  \centering
  \includegraphics{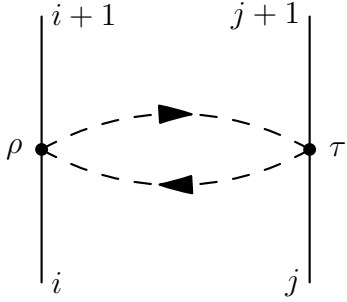}
  \caption{The two-scalar exchange contribution to the super-Wilson loop.  It is of order four in the \(\theta\) expansion.}
  \label{fig:ss}
\end{figure}

Let us now discuss the answer we obtain from the correlation functions
involving scalar bilinears (see fig.~\ref{fig:ss}). The relevant terms
we are interested in appear in the odd part of the connection and read
\begin{equation}
  - \frac i g e_{F}^{K L \beta} \bar{\theta}_{M K \beta} \left(\phi^{M}
  \bar{\phi}_{L} - \frac 1 4 \delta_{L}^{M} \phi^{P} \bar{\phi}_{P}\right),
\end{equation} where \(e_{F}\) is the fermionic vielbein \(e_{F}^{K L
  \beta} = d \theta^{K L \beta}\).  If we restrict this to a
light-like line between points \((x_{i}, \theta_{i})\) and \((x_{i+1},
\theta_{i+1})\), with parametrization \(\theta(\tau) = \theta_{i} + \tau
(\theta_{i+1}-\theta_{i})\), we obtain \(e_{F}^{K L \beta} = -d \tau
\lambda_{i+1}^{\beta} \eta_{i+1}^{L N}\).  When dotted into
\(\bar{\theta}(\tau)\) the factor \(\lambda_{i+1}\) makes the
dependence on \(\tau\) disappear and we are left with
\begin{equation}
  d \tau \bar{\theta}_{i,MK\beta} \lambda_{i+1}^{\beta} \eta_{i+1}^{K
    L} = d \tau \tilde{\chi}_{i+1,MK} \eta_{i+1}^{K L}.
\end{equation}

Now we take two sides between vertices \((i,i+1)\) and \((j,j+1)\) and
compute the correlation function between the bosonic bilinears.  The
position on the first line is parametrized by \(\tau\) and on the
second line by \(\sigma\).  If we label the position at which the fields
are evaluated by \(\tau\) and \(\sigma\) we find
\begin{equation}
  d \tau \tilde{\chi}_{i+1,MK} \eta_{i+1}^{KN} d \sigma
  \tilde{\chi}_{j+1,PL} \eta_{i+1}^{LQ} \frac {\delta_{Q}^{M}
    \delta_{N}^{P} - \frac 1 4 \delta_{M}^{N}
    \delta_{P}^{Q}}{(x(\tau)-x(\sigma))^{2}}.
\end{equation}  The \(\eta\) can be expanded in terms of \(\chi\) so
everything can be written in terms of twistor components.

This formula produces an unusual contraction pattern of the \(SU(4)\)
indices \(\tilde{\chi}_{i+1,MK} \eta_{i+1}^{KN} \tilde{\chi}_{j+1,N L}
\eta_{i+1}^{LM}\).  Superconformal invariants are built out of
products \(Z_{i} \cdot \bar{Z}_{j}\) whose nilpotent part is
\(\chi_{i,MN} \tilde{\chi}_{j}^{MN}\).  At first sight it looks like
the combination above can not be written in this form.  Nevertheless,
we will show that it is possible to rewrite it in such a form.  For
this we consider two Grassmann variables \(\psi_{MN}\),
\(\chi_{PQ}\) which are antisymmetric in the exchange \(M
\leftrightarrow N\) and \(P \leftrightarrow Q\).  Then, we have the
following identity
\begin{multline}
  \psi_{MN} \chi_{PQ} - \psi_{MP} \chi_{NQ} + \psi_{MQ} \chi_{NP} +
  \psi_{NP} \chi_{MQ} -\\ \psi_{NQ} \chi_{MP} + \psi_{PQ} \chi_{MN} =
  \frac 1 4 \epsilon_{MNPQ} \epsilon^{RSTU} \psi_{RS} \chi_{TU}.
\end{multline}  This identity can be proved by noticing that the
left-hand side is completely antisymmetric in \(MNPQ\) and therefore
it is proportional to \(\epsilon_{MNPQ}\).  The proportionality
constant can be obtained by contracting with \(\epsilon^{MNPQ}\).

Using this identity we can show that, given two other Grassmann
variables \(A\) and \(B\) which are also antisymmetric in their two
\(SU(4)\) indices, we have
\begin{equation}
  \label{eq:ferm-rearrangement}
  (\psi A \chi B) - (\psi B)(A \chi) - (\psi A)(\chi B) = -
  (\psi \chi)  (A B),
\end{equation} where \(\psi A \chi B = \psi_{IJ} A^{JK} \chi_{KL}
B^{LI}\) and, as always, we normalize the index contractions such that
\(\psi A = \tfrac 1 2 \psi_{IJ} A^{IJ}\).  This formula allows us to
rewrite the final answer in terms of objects whose origin is more
clear from the point of view of superconformal symmetry.

\section{Perturbative expansion vs.\ Grassmann expansion}
\label{sec:expansion}

We now want to understand how to compute the perturbative order of a given
planar diagram contributing to the expectation value of a Wilson loop. In
particular, we would like to relate the perturbative order with the topology of
the diagram.

%
The most convenient way to explicit this relation is by rescaling the matter
fields by a factor of $\sqrt{g}$. This way, all the interaction vertices
contribute a factor of $g$ and all propagators a factor of $g^{-1}$. Most
importantly, with this rescaling the superconnection does not contribute any
power of $g$, since this rescaling cancels the $g^{-1}$ dependence in the
Grassmann expansion. The perturbative order $j$ associated with a given graph is
then easily expressed in terms of the total number of propagators $P$ and
interaction vertices $V_{int}$ as
\begin{equation}
  \label{eq:pv}
  j = V_{int} - P
\end{equation}

We are interested in polygonal Wilson loops in the planar limit, thus homotopic
to a disk. The Euler characteristic is $1$, therefore the total number of
vertices $V$, of edges $E$ and of faces $F$ satisfy
\begin{equation}
  \label{eq:euler} V-E+F = 1
\end{equation}
The total number of vertices is given by the number of interaction vertices plus
the number of operator insertions on the boundary, $V=V_{int} +
V_{\partial}$. Similarly, the number of edges is equal to the number of
propagators plus the number of segments in which the boundary is partitioned by
the insertions, $E=P+E_\partial$. Obviously, $V_\partial = E_\partial$,
therefore the Euler characteristic equation becomes
\begin{equation}
  \label{eq:pertord}
  V_{int} - P = 1 - F
\end{equation}
In the right hand side one can recognize the perturbative order $j$. Therefore,
given a graph with $F$ faces, it contributes at order $g^{1-F}$.

The \(\theta\) expansion of the gauge connections can be done
explicitly to arbitrarily high order.  Here we only want to comment on
some of the qualitative features of this expansion.  We have that a
term of schematic form \(\nabla^{k} \phi^{l} \psi^{m} \in A_{F}\) with
\(l+m\) even and \(l+m \geq 2\) contributes to orders \(\theta^{2 k +
  l + 2 m - 1}\) and \(g^{-(l+m)/2}\).  Terms of schematic type
\(\nabla^{n} \phi^{p} \psi^{q} \in A_{B}\) with \(p+q\) even and
\(p+q\geq 2\) contribute to orders \(\theta^{2 n + p + 2 q - 2}\) and
\(g^{-(p+q)/2}\).  The types of fields resulting from this double
expansion are listed in
tables~\ref{tab:AF-expansion}~\ref{tab:AB-expansion}.

\begin{table}
  \centering
  \begin{tabular}{|l|*{6}{c|}}\hline
    \backslashbox{\(g^{\#}\)}{\(\theta^{\#}\)}
      & \(0\) & \(1\) & \(2\) & \(3\) & \(4\) & \(5\)\\
    \hline
    \(0\) & \(0\) & \(a\) & \(0\) & \(0\) & \(0\) & \(0\)\\
    \hline
    \(-1\) & \(0\) & \(\phi^{2}\) & \(\phi \psi\) & \(\nabla
    \phi^{2}\), \(\psi^{2}\) & \(\nabla \phi \psi\) & \(\nabla^{2}
    \phi^{2}\), \(\nabla \psi^{2}\)\\
    \hline
    \(-2\) & \(0\) & \(0\) & \(0\) & \(\phi^{4}\) & \(\phi^{3} \psi\)
    & \(\nabla \phi^{4}\)\\
    \hline
    \(-3\) & \(0\) & \(0\) & \(0\) & \(0\) & \(0\) & \(\phi^{6}\),
    \(\phi^{2} \psi^{2}\)\\
    \hline
  \end{tabular}
  \caption{The interplay between the \(\theta\) expansion and the expansion in the coupling for the fermionic gauge connection.}
  \label{tab:AF-expansion}
\end{table}

\begin{table}
  \centering
  \begin{tabular}{|l|*{6}{c|}}\hline
    \backslashbox{\(g^{\#}\)}{\(\theta^{\#}\)}
      & \(0\) & \(1\) & \(2\) & \(3\) & \(4\) & \(5\)\\
    \hline
    \(0\) & \(a\) & \(0\) & \(0\) & \(0\) & \(0\) & \(0\)\\
    \hline
    \(-1\) & \(0\) & \(\phi \psi\) & \(\nabla \phi^{2}\), \(\psi^{2}\)
    & \(\nabla \phi \psi\) & \(\nabla^{2} \phi^{2}\), \(\nabla
    \psi^{2}\) & \(\nabla^{2} \phi \psi\)\\
    \hline
    \(-2\) & \(0\) & \(0\) & \(\phi^{4}\) & \(\phi^{3} \psi\) &
    \(\phi^{2} \psi^{2}\), \(\nabla \phi^{4}\) & \(\nabla \phi^{3}
    \psi\), \(\phi \psi^{3}\)\\
    \hline
    \(-3\) & \(0\) & \(0\) & \(0\) & \(0\) & \(\phi^{6}\), \(\phi^{2}
    \psi^{2}\) & \(\psi \phi^{5}\)\\
    \hline
  \end{tabular}
  \caption{The interplay between the \(\theta\) expansion and the expansion in the coupling for the bosonic gauge connection.}
  \label{tab:AB-expansion}
\end{table}

\section{A peek at three loops}
\label{sec:3loop}

\begin{figure}
  \centering
  \begin{subfigure}[b]{0.3\textwidth}
    \includegraphics[width=\textwidth]{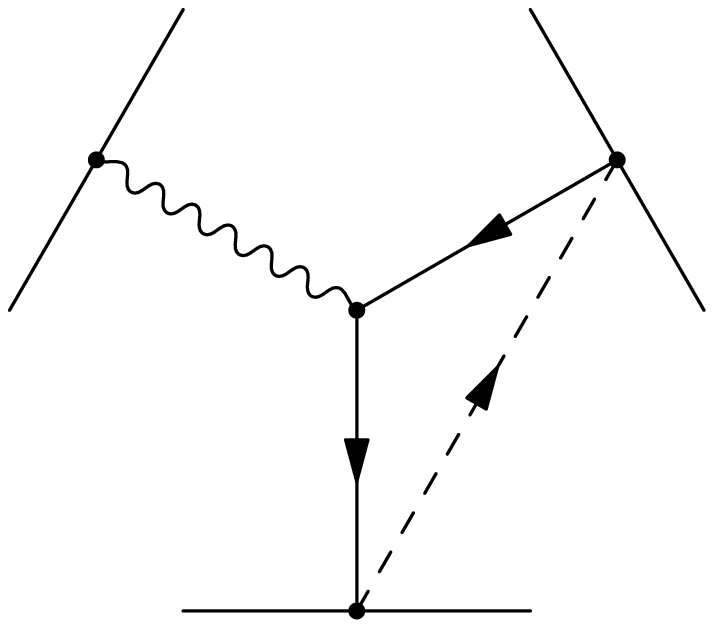}
    \label{fig:g3wt2A}
  \end{subfigure}%
  \quad
  \begin{subfigure}[b]{0.3\textwidth}
    \includegraphics[width=\textwidth]{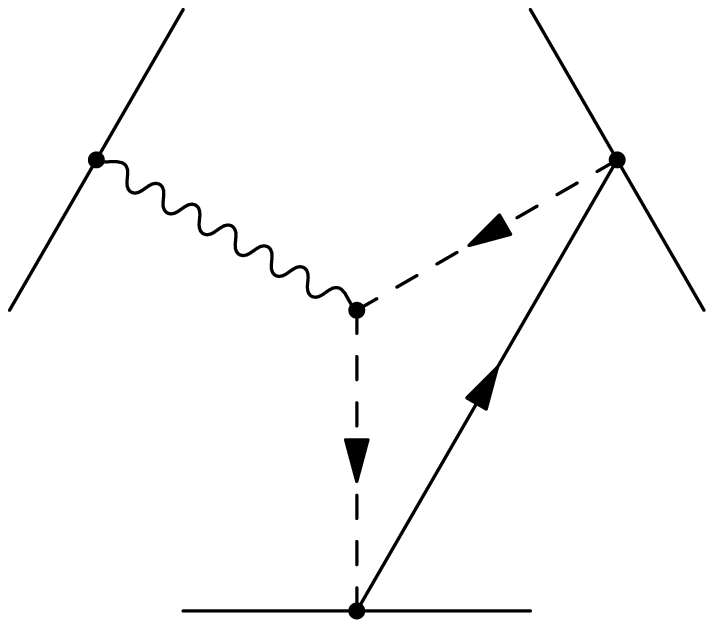}
    \label{fig:g3wt2B}
  \end{subfigure}%
  \quad
  \begin{subfigure}[b]{0.3\textwidth}
    \includegraphics[width=\textwidth]{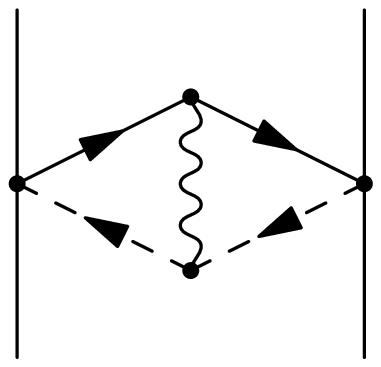}
    \label{fig:g3wt2C}
  \end{subfigure}
  \begin{subfigure}[b]{0.3\textwidth}
    \includegraphics[width=\textwidth]{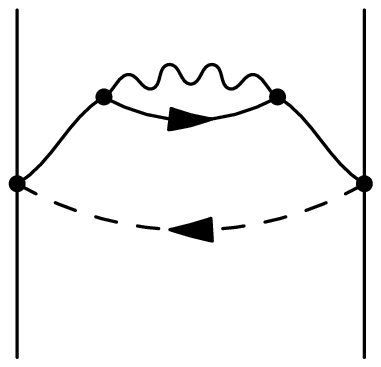}
    \label{fig:g3wt2D}
  \end{subfigure}%
  \quad
  \begin{subfigure}[b]{0.3\textwidth}
    \includegraphics[width=\textwidth]{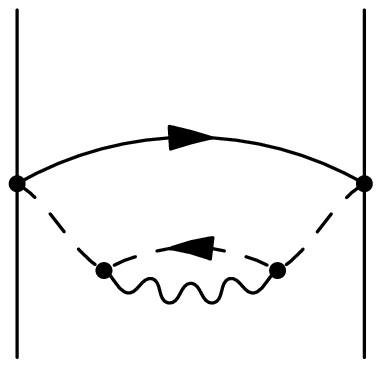}
    \label{fig:g3wt2E}
  \end{subfigure}%
  \quad
  \begin{subfigure}[b]{0.3\textwidth}
    \includegraphics[width=\textwidth]{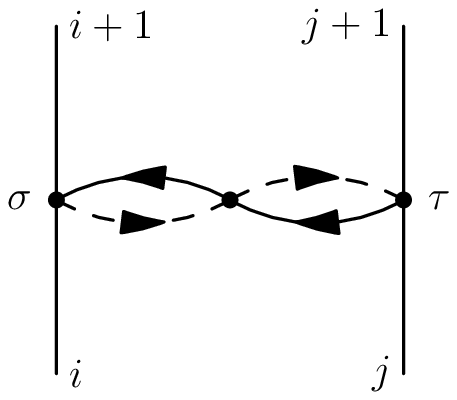}
    \label{fig:g3wt2F}
  \end{subfigure}%
  \caption{The diagrams contributing to order \(g^{-3}\) and to Grassmann weight two.}
  \label{fig:g3wt2}
\end{figure}

\begin{figure}
  \centering
  \includegraphics{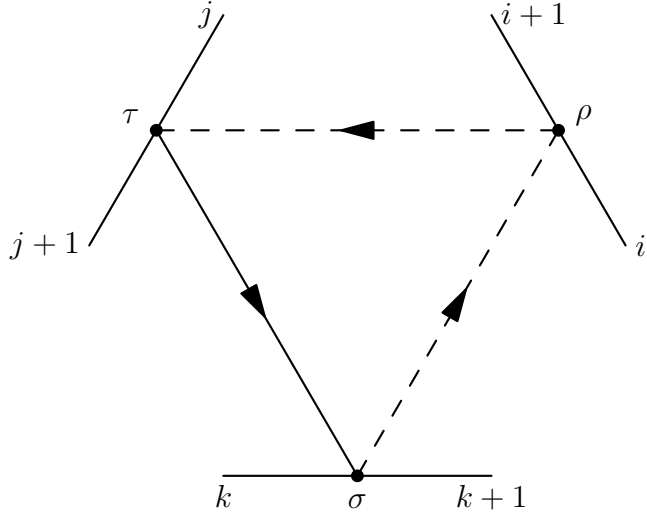}
  \caption{A diagram contributing to order \(g^{-3}\) and to Grassmann
    weight four.  There are some other possibilities involving two
    scalar exchanges interacting with photons which can be obtained
    by replacing the fermion line in fig.~\ref{fig:g3wt2} by a scalar
    line.  However, if the \(i\), \(j\) and \(k\) sides of the Wilson
    loop are separated by one or more sides, such diagrams do not
    contribute the same kind of Grassmannian quartic terms.}
  \label{fig:g3wt4}
\end{figure}

Let us now consider the three-loop, i.e.\ \(g^{-3}\) contributions to
the super-Wilson loop.  Among these contributions, we can look at
terms with Grassmann weight zero, two, four, etc.  We will see that
the complexity of the computation decreases as we increase the
Grassmann weight.  For example, in fig.~\ref{fig:g3wt2} we show the
diagrams which contribute to the Grassmann weight two part and in
fig~\ref{fig:g3wt4} we show the diagram which contribute to the
Grassmann weight four part.  Needless to say, the Grassmann order zero
part is a sum of many more diagrams.

For the contribution of the diagram in fig.~\ref{fig:g3wt4} we obtain
after a fermionic rearrangement as in
eq.~\eqref{eq:ferm-rearrangement} and partial translation to twistor
language
\begin{equation}
  \label{eq:g3wt4}
  \frac {z_{j+1} \cdot \tilde{z}_{k+1}}{2 (16 \pi g)^{3}} \int_{0}^{1}
  d \rho \int_{0}^{1} d \tau \int_{0}^{1} d \sigma
  \frac{(\chi_{i+1} \cdot \tilde{\chi}_{j+1}) (\eta_{i+1} \cdot
    \tilde{\chi}_{k+1}) - (j \leftrightarrow k)}{\lvert x_{i}(\rho) -
    x_{j}(\tau)\rvert \lvert x_{i}(\rho) - x_{k}(\sigma)\rvert \lvert
    x_{j}(\tau) - x_{k}(\sigma)\rvert^{3}},
\end{equation} where \(x_{i}(\rho) = (1-\rho) x_{i} + \rho x_{i+1}\),
etc.

Using the expression for \(\eta_{i+1}\) we see that this result
contains several types of quartic terms in \(\chi\).  The terms of
type \((\chi_{i+1} \cdot \tilde{\chi}_{j+1}) (\chi_{i+1} \cdot
\tilde{\chi}_{k+1})\) cancel by antisymmetry in variables \(j\) and
\(k\).  Up to symmetries, the only other type of entry is \((\chi_{i}
\cdot \tilde{\chi}_{j+1}) (\chi_{i+1} \cdot \tilde{\chi}_{k+1})\).  It
appears in a combination
\begin{multline}
  \frac {z_{j+1} \cdot \tilde{z}_{k+1}}{2 (16 \pi g)^{3} \langle i
    i+1\rangle} (\chi_{i} \cdot \tilde{\chi}_{j+1}) (\chi_{i+1} \cdot
\tilde{\chi}_{k+1}) \int_{0}^{1} d \rho \int_{0}^{1} d \tau \int_{0}^{1} d
  \sigma \frac 1 {\lvert x_{j}(\tau) - x_{k}(\sigma)\rvert^{3}} \times\\
  \left(\frac 1 {\lvert x_{i-1}(\rho) - x_{j}(\sigma)\rvert \lvert
      x_{i-1}(\rho) - x_{k}(\tau)\rvert} -\frac 1 {\lvert x_{i}(\rho) - x_{j}(\sigma)\rvert \lvert x_{i}(\rho) - x_{k}(\tau)\rvert} \right).
\end{multline}

The integrals above are of the form
\begin{equation}
  \label{eq:int-type}
  \int_{0}^{1} d \rho \int_{0}^{1} d \tau \int_{0}^{1} d \sigma
  P_{1}(\tau, \rho)^{-1/2} P_{2}(\rho, \sigma)^{-1/2} P_{3}(\sigma,
  \tau)^{-3/2},
\end{equation} where \(P_{i}(x,y) = A_{i} x y + B_{i} x + C_{i} y +
D_{i}\) and \(A_{i}, B_{i}, C_{i}, D_{i}\) are some constants.  Two of
these integrals can be performed using the formulas
\begin{gather}
  \int_{0}^{1} \frac {d t}{\sqrt{(m t + n)(p t + q)^{3}}} = - \frac 2
  {n p-m q} \left(\frac {\sqrt{m+n}}{\sqrt{p+q}} - \frac
    {\sqrt{n}}{\sqrt{q}}\right),\\
  \int_{0}^{1} \frac {d t}{\sqrt{a t+b} (c t+d) \sqrt{e t+f}} = \frac
  1 {\sqrt{(a d-b c)(d e-c f)}} \Biggl[\ln (1+\frac c d) -\\ 2 \ln
  \Bigl(\frac {\sqrt{a d-b c} \sqrt{e+f} - \sqrt{d e-c f} \sqrt{a+b}}{\sqrt{a d-b c} \sqrt{f} - \sqrt{d e-c f} \sqrt{b}}\Bigr)\Biggr].
\end{gather} At this stage, it appears unlikely that the third integral in
eq.~\eqref{eq:int-type} can be computed in terms of classical
polylogarithms.

In general, when integrating expressions of type \(\int d t R(t) \ln
S(t)\) where \(R\) and \(S\) are rational fractions in \(t\) the
result can be expressed in terms of dilogarithms.  However, here we
have more complicated expressions due to the presence of square
roots.

It is noteworthy that when performing the integrals as described above
we obtain the square root of a cubic polynomial in the last
integration step.  Is this a hint that the result is an elliptic
polylogarithm?  For arbitrary values of \(A_{i}, B_{i}, C_{i}, D_{i}\)
this cubic polynomial is generic, but for six points \(x_{i}, x_{i+1},
x_{j}, x_{j+1}, x_{k}, x_{k+1}\) there is a Gram determinant
constraint which imposes a constraint among the \(12\) values \(A_{i},
B_{i}, C_{i}, D_{i}\).  It is curious that when this constraint is
satisfied the cubic polynomial mentioned above factorizes into a
linear polynomial and the square of another linear polynomial.

Another way to understand this constraint is to think more deeply
about the geometry of the problem.  We have three lines containing
points \((x_{i}, x_{i+1})\), \((x_{j}, x_{j+1})\) and \((x_{k},
x_{k+1})\) respectively.  Let us now find the transversals, i.e.\ the
light-like lines intersecting all of three lines (an analogous problem
in 4D was considered in refs.~\cite{Hodges2010, Arkani-Hamed2010}).
To find a light-like line intersecting lines \((x_{i}, x_{i+1})\),
\((x_{j}, x_{j+1})\) and \((x_{k}, x_{k+1})\) we pick a point on each
one of them with parameter \(\rho\), \(\sigma\) and \(\tau\)
respectively.  Then, the light-like conditions read \(P_{1}(\tau,
\rho) = 0\), \(P_{2}(\rho, \sigma) = 0\) and \(P_{3}(\sigma, \tau) =
0\).  The first two equations are linear in \(\tau\) and \(\sigma\) so
we can trivially solve for them and plug back in the third.  We obtain
a degree two polynomial whose discriminant is zero since it is equal
to the Gram determinant constraint for six points in three dimensions.

Therefore, in three dimensions, there is a unique (with multiplicity
two) light-like line intersecting any three non-intersecting
light-like lines.  If we denote by \(\rho_{*}\), \(\sigma_{*}\) and
\(\tau_{*}\) the parameters of the intersection points and perform a
change of variables \(r = \rho - \rho_{*}\), \(s = \sigma -
\sigma_{*}\) and \(t = \tau - \tau_{*}\) and we also rescale the
polynomials \(P_{i}\) such as to make the leading coefficient equal to
unity, then we reduce the problem to performing the integral
\begin{equation}
  \int_{-\rho_{*}}^{1-\rho_{*}} d r \int_{-\sigma_{*}}^{1-\sigma_{*}}
  d s \int_{-\tau_{*}}^{1-\tau_{*}} d t Q_{1}(t,r)^{-1/2}
  Q_{2}(r,s)^{-1/2} Q_{3}(s,t)^{-3/2},
\end{equation} where \(Q_i(x,y) = x y + b_{i} x + c_{i} y\) and such
that \(b_{1} b_{2} b_{3} + c_{1} c_{2} c_{3} = 0\).

The integral in eq.~\eqref{eq:int-type} can be transformed by using the
star-triangle identitites (see ref.~\cite{Ussyukina1991382}). The triangle in
eq.~\eqref{eq:int-type} is a semi-unique triangle in the language of this
reference. It can be computed in terms of triangles with exponents
$\tfrac{3}{2}, \tfrac{3}{2}, -\tfrac{1}{2}$ and of stars with exponents
$2,1,1$. The triangles with those exponents can be computed in terms of
dilogarithms. In the star integral all of the fractional powers disappear, and
the integrals over $\rho,\tau$ and $\sigma$ factorize and can be computed
straightforwardly; however, the remaining integrals over the vertex of the star
involve the square of a logarithm times a rational function.  These integrals
seem to be just as complicated as the original ones.

It is also worth mentioning that the integral in eq.~\eqref{eq:int-type}
simplifies at low number of points, even though it remains still quite
complicated. At low number of points, it is possible to translate some of the
vertices to \(\theta=0\); due to this fact, supersymmetry is more powerful at
higher number of points, therefore we focus on these cases.

We can try to compute the integral in fig.~\ref{fig:g3wt4} in special
kinematics.  For example, if the sides \((i,i+1)\), \((j,j+1)\) and
\((k,k+1)\) of the Wilson loop belong to the same two-dimensional
plane the kinematics simplifies enough to allow an explicit evaluation
of the integral and the answer is rational.  This is not in
contradiction with the fact that the \(g^{-3}\) part of the answer is
of transcendentality three.  Indeed, consider the dilogarithm
\(\li_{2}(x)\) whose derivative is \(-\frac {\ln(1-x)}{x}\), whose limit is equal
to \(1\) when evaluated at \(x=0\).  Another example which is closer
to the one we are considering here is to take two derivatives and one
limit.  If we do this for \(\li_{3}(x)\) and we compute the limit \(x
\to 0\) we again obtain a rational answer.

\acknowledgments

We thank Niklas Beisert and Burkhard Schwab for discussions.
The work of MR is partially supported
by grant no.\ 200021-137616 from the Swiss National Science Foundation.

\bibliographystyle{JHEP}
\bibliography{swl}

\end{document}